\documentclass[sigconf,nonacm]{acmart}
\pdfoutput=1

\usepackage{amsmath,mathtools}
\usepackage{booktabs}
\usepackage{enumitem}
\usepackage{hyperref}
\usepackage[capitalise,noabbrev]{cleveref}
\usepackage{tikz}
\usepackage{pgfplots}
\pgfplotsset{compat=1.18}
\usetikzlibrary{arrows.meta,positioning,calc,decorations.pathreplacing,patterns,shapes.geometric}

\definecolor{NavyBlue}{RGB}{24,78,158}
\definecolor{RoyalBlue}{RGB}{45,112,205}
\definecolor{BrickRed}{RGB}{188,43,38}
\definecolor{ForestGreen}{RGB}{26,128,68}
\definecolor{Orange}{RGB}{217,112,14}
\definecolor{Goldenrod}{RGB}{242,196,65}

\renewcommand\footnotetextcopyrightpermission[1]{}
\AtBeginDocument{%
  \setlength{\abovedisplayskip}{6pt plus 6pt minus 2pt}%
  \setlength{\belowdisplayskip}{6pt plus 6pt minus 2pt}%
}

\title{The Anatomy of Piecewise-Linear Learned Indexes}

\author{Yue Chen}
\affiliation{%
  \institution{Department of Computer Science, Tsinghua University}
  \city{Beijing}
  \country{China}
}
\email{yue-chen25@mails.tsinghua.edu.cn}

\author{Yong Zhang}
\affiliation{%
  \institution{Beijing National Research Center for Information Science and Technology, Tsinghua University}
  \city{Beijing}
  \country{China}
}
\email{zhangyong05@tsinghua.edu.cn}

\newtheorem{theorem}{Theorem}[section]
\newtheorem{lemma}[theorem]{Lemma}
\newtheorem{proposition}[theorem]{Proposition}
\newtheorem{corollary}[theorem]{Corollary}
\theoremstyle{definition}

\theoremstyle{remark}
\newtheorem{remark}[theorem]{Remark}
\AddToHook{env/lemma/begin}{\crefalias{theorem}{lemma}}
\AddToHook{env/proposition/begin}{\crefalias{theorem}{proposition}}
\AddToHook{env/corollary/begin}{\crefalias{theorem}{corollary}}
\AddToHook{env/definition/begin}{\crefalias{theorem}{definition}}
\AddToHook{env/remark/begin}{\crefalias{theorem}{remark}}
\AddToHook{cmd/appendix/after}{\crefalias{section}{appendix}}

\newcommand{\R}{\mathbb{R}}
\newcommand{\N}{\mathbb{N}}
\newcommand{\E}{\mathbb{E}}
\newcommand{\Prob}{\mathbb{P}}
\newcommand{\Var}{\mathrm{Var}}

\newcommand{\eps}{\varepsilon}
\newcommand{\iid}{i.i.d.}
\newcommand{\kopt}{\kappa_{\mathrm{opt}}}
\newcommand{\aopt}{\alpha_{\mathrm{opt}}}
\newcommand{\Cinf}{C_{\infty}}
\newcommand{\Ceps}{C_{\eps}}
\newcommand{\tauopt}{\tau_{\mathrm{opt}}}
\newcommand{\tauan}{\tau_{\mathrm{an}}}

\newcommand{\kan}{\kappa_{\mathrm{an}}}
\newcommand{\kch}{\kappa_{\mathrm{ch}}}

\newcommand{\rhostar}{\rho_*}
\newcommand{\Fclass}{\mathfrak{F}}
\newcommand{\abs}[1]{\left|#1\right|}

\newcommand{\ceil}[1]{\left\lceil#1\right\rceil}
\newcommand{\floor}[1]{\left\lfloor#1\right\rfloor}
\DeclareUrlCommand\leanid{\urlstyle{tt}}

\begin{document}

\begin{abstract}
Piecewise-linear approximation (PLA) learned indexes---including the PGM-index, FITing-Tree, RadixSpline, PLEX, and PLIN---compress a sorted key array into error-bounded line segments and route queries through an upper directory, yet prior theory stops at coarse $\Theta(\eps^2)$ bounds on a single segment.
At the bottom layer, we prove that optimal greedy PLA covers $\kopt\eps^2/v_0 \approx 3.54\,\eps^2/v_0$ keys per segment under independent gaps and $\kopt\eps^2$ under any smooth density $f$ ($v_0 = 1$), while one-pass anchored and chord fitters preserve $\Theta(\eps^2/v_0)$ coverage.
When PLA layers are stacked recursively, a law-of-large-numbers argument---recently invoked to claim $O(\log\log n)$ height for the standard PGM-index---suggests that summing $\Theta(\eps^2)$ gaps per segment should average out gap fluctuations across layers.
We prove that this argument fails under standard Endpoint Promotion (EP): random stopping-time fluctuations reset the relative gap variance at every depth, locking recursive fanout at a universal constant $\Ceps \approx 19.861\,\eps^2$ and trapping expected height at $\log n / \log \Ceps + O(1)$ and lookup cost at $\Theta(\log n)$.

To break this logarithmic barrier, we introduce Rank-Equispaced Promotion (RE), which samples promoted pivots at a uniform rank stride.
Each RE stride sums a fixed number of base gaps rather than a random stopping-time count, so relative gap variance decays across layers and cumulative segment coverage squares at every step.
Expected height drops to $\log_2\log_\eps n + O(1)$ on independent gaps and $\log_2\log_\eps n + \log_2^+(\log n/\log(\eps/\delta)) + O(1)$ on smooth densities with curvature $\delta$.
A compact bucket directory routes queries across unaligned RE pivots with fewer than one extra comparison per layer on average, giving RE-PGM $O(\log\eps \cdot \log\log_\eps n)$ expected lookup in $\Theta(n/\eps^2)$ space and $O(n)$ build time while avoiding the key-space skew sensitivity of flat radix tables.
Our exact constants predict empirical fanout within $0.3\%$ down to $\eps = 2$, and all core theorems are machine-checked in Lean~4.
\end{abstract}

\maketitle

\section{Introduction}\label{sec:intro}

\begin{figure*}[t]
\centering
\begin{tikzpicture}[
    x=1cm, y=1cm,
    >={Stealth[length=4.2pt, width=3.2pt]},
    smallarr/.style={{Stealth[length=3.0pt, width=2.3pt]}-{Stealth[length=3.0pt, width=2.3pt]}, thin},
    font=\footnotesize,
    paneltitle/.style={font=\footnotesize\bfseries, text=black!90, anchor=west},
    axisline/.style={-{Stealth[length=4.5pt, width=3.4pt]}, draw=black!75, semithick},
    fitline/.style={draw=NavyBlue!90!black, very thick},
    boundline/.style={draw=NavyBlue!55, densely dashed, semithick},
    keypt/.style={circle, fill=NavyBlue!85!black, inner sep=1.4pt},
    rejectpt/.style={circle, fill=BrickRed!90!black, draw=white, semithick, inner sep=1.8pt},
    eppt/.style={circle, fill=BrickRed!85!black, draw=white, thin, inner sep=1.65pt},
    rept/.style={regular polygon, regular polygon sides=4, fill=ForestGreen!80!black, draw=white, thin, inner sep=1.45pt}
]

\draw[black!15, thin] (5.55, 0.10) -- (5.55, 4.15);
\draw[black!15, thin] (11.85, 0.10) -- (11.85, 4.15);

\node[paneltitle] at (0.05, 4.02) {(a) Greedy $\pm\eps$ Corridor};

\draw[axisline] (0.55, 0.50) -- (5.15, 0.50);
\node[font=\scriptsize, text=black!85, anchor=north east] at (5.15, 0.42) {key $x$};
\draw[axisline] (0.55, 0.50) -- (0.55, 3.65);
\node[font=\scriptsize, text=black!85, anchor=north west] at (0.62, 3.65) {rank $i$};

\fill[NavyBlue!11] (0.78, 0.58) -- (2.95, 1.665) -- (2.95, 2.265) -- (0.78, 1.18) -- cycle;
\draw[boundline] (0.78, 1.18) -- (2.95, 2.265);
\draw[boundline] (0.78, 0.58) -- (2.95, 1.665);
\draw[fitline] (0.78, 0.88) -- (2.95, 1.965);

\draw[NavyBlue!45, densely dotted, semithick] (2.95, 1.965) -- (3.42, 2.20);
\draw[NavyBlue!45, densely dotted, semithick] (2.95, 2.265) -- (3.42, 2.50);

\draw[{Stealth[length=2.5pt, width=2.0pt]}-{Stealth[length=2.5pt, width=2.0pt]}, NavyBlue!80!black, thin]
    (1.24, 1.12) -- (1.24, 1.40);
\node[font=\scriptsize\bfseries, text=NavyBlue!90!black, anchor=south east, inner sep=1pt] at (1.22, 1.43) {$\pm\eps$};

\node[keypt] at (0.82, 0.80) {};
\node[keypt] at (1.35, 1.04) {};
\node[keypt] at (1.82, 1.54) {};
\node[keypt] at (2.38, 1.58) {};
\node[keypt] at (2.88, 2.04) {};
\node[font=\scriptsize\bfseries, text=NavyBlue!90!black, anchor=north west] at (1.85, 1.12) {$S_1$};

\fill[RoyalBlue!11] (3.32, 2.42) -- (4.95, 3.105) -- (4.95, 3.665) -- (3.32, 2.98) -- cycle;
\draw[boundline] (3.32, 2.98) -- (4.95, 3.665);
\draw[boundline] (3.32, 2.42) -- (4.95, 3.105);
\draw[fitline] (3.32, 2.70) -- (4.95, 3.385);
\node[font=\scriptsize\bfseries, text=NavyBlue!90!black, anchor=north west] at (4.05, 2.74) {$S_2$};

\node[keypt] at (3.78, 2.98) {};
\node[keypt] at (4.32, 3.02) {};
\node[keypt] at (4.82, 3.44) {};

\draw[{Stealth[length=3.0pt, width=2.2pt]}-{Stealth[length=3.0pt, width=2.2pt]}, BrickRed!90!black, semithick]
    (3.32, 2.16) -- (3.32, 2.71);
\node[font=\scriptsize\bfseries, text=BrickRed!90!black, anchor=east] at (3.26, 2.44) {$>\!\eps$};
\node[rejectpt] at (3.32, 2.78) {};
\node[font=\scriptsize\bfseries, text=BrickRed!90!black, anchor=south east] at (3.28, 2.84) {$p_5$};

\draw[black!55, thin] (0.78, 0.14) -- (0.78, 0.28);
\draw[black!55, thin] (2.95, 0.14) -- (2.95, 0.28);
\draw[smallarr, black!70] (0.78, 0.21) -- (2.95, 0.21)
    node[midway, fill=white, inner sep=1.5pt, font=\scriptsize, text=black!85] {$T_1 = 5$};

\node[paneltitle] at (5.80, 4.02) {(b) EP vs.\ RE Promotion};

\node[font=\scriptsize\bfseries, text=NavyBlue!90!black, anchor=west] at (5.85, 3.22) {PLA};
\node[font=\scriptsize\bfseries, text=BrickRed!85!black, anchor=west] at (5.85, 2.08) {EP};
\node[font=\scriptsize\bfseries, text=ForestGreen!80!black, anchor=west] at (5.85, 0.86) {RE};

\draw[fill=NavyBlue!12, draw=NavyBlue!75, semithick, rounded corners=1.5pt] (6.75, 3.02) rectangle (7.97, 3.42);
\node[font=\scriptsize\bfseries, text=NavyBlue!90!black] at (7.36, 3.22) {$S_1$};

\draw[fill=NavyBlue!12, draw=NavyBlue!75, semithick, rounded corners=1.5pt] (8.03, 3.02) rectangle (9.47, 3.42);
\node[font=\scriptsize\bfseries, text=NavyBlue!90!black] at (8.75, 3.22) {$S_2$};

\draw[fill=NavyBlue!12, draw=NavyBlue!75, semithick, rounded corners=1.5pt] (9.53, 3.02) rectangle (10.72, 3.42);
\node[font=\scriptsize\bfseries, text=NavyBlue!90!black] at (10.125, 3.22) {$S_3$};

\draw[fill=NavyBlue!12, draw=NavyBlue!75, semithick, rounded corners=1.5pt] (10.78, 3.02) rectangle (11.50, 3.42);
\node[font=\scriptsize\bfseries, text=NavyBlue!90!black] at (11.14, 3.22) {$S_4$};

\foreach \r in {0, 5, 11, 16} {
    \pgfmathsetmacro{\xpos}{6.75 + \r * 0.250}
    \draw[BrickRed!45, densely dashed, thin] (\xpos, 3.00) -- (\xpos, 2.38);
}

\draw[black!45, semithick] (6.75, 2.08) -- (11.50, 2.08);
\foreach \r/\lbl in {0/p_0, 5/p_5, 11/p_{11}, 16/p_{16}} {
    \pgfmathsetmacro{\xpos}{6.75 + \r * 0.250}
    \node[eppt] at (\xpos, 2.08) {};
    \node[font=\scriptsize, text=BrickRed!90!black, above=2pt] at (\xpos, 2.08) {$\lbl$};
}
\draw[smallarr, BrickRed!80!black] (6.78, 1.72) -- (7.97, 1.72)
    node[midway, fill=white, inner sep=1pt, font=\tiny\bfseries, text=BrickRed!85!black] {$5$};
\draw[smallarr, BrickRed!80!black] (8.03, 1.72) -- (9.47, 1.72)
    node[midway, fill=white, inner sep=1pt, font=\tiny\bfseries, text=BrickRed!85!black] {$6$};
\draw[smallarr, BrickRed!80!black] (9.53, 1.72) -- (10.72, 1.72)
    node[midway, fill=white, inner sep=1pt, font=\tiny\bfseries, text=BrickRed!85!black] {$5$};
\draw[smallarr, BrickRed!80!black] (10.78, 1.72) -- (11.50, 1.72)
    node[midway, fill=white, inner sep=1pt, font=\tiny\bfseries, text=BrickRed!85!black] {$3$};

\draw[black!45, semithick] (6.75, 0.86) -- (11.50, 0.86);
\fill[Goldenrod!32, draw=Orange!85!black, semithick, rounded corners=1pt] (8.75, 0.73) rectangle (9.75, 0.99);
\draw[BrickRed!90!black, thick] (9.50, 0.69) -- (9.50, 1.03);

\foreach \r/\lbl in {0/p_0, 4/p_4, 8/p_8, 12/p_{12}, 16/p_{16}} {
    \pgfmathsetmacro{\xpos}{6.75 + \r * 0.250}
    \node[rept] at (\xpos, 0.86) {};
    \node[font=\scriptsize, text=ForestGreen!85!black, above=2pt] at (\xpos, 0.86) {$\lbl$};
}
\foreach \xa/\xb in {6.78/7.72, 7.78/8.72, 8.78/9.72, 9.78/10.72} {
    \draw[smallarr, ForestGreen!80!black] (\xa, 0.50) -- (\xb, 0.50)
        node[midway, fill=white, inner sep=1pt, font=\tiny\bfseries, text=ForestGreen!85!black] {$4$};
}
\draw[smallarr, ForestGreen!80!black] (10.78, 0.50) -- (11.50, 0.50)
    node[midway, fill=white, inner sep=1pt, font=\tiny\bfseries, text=ForestGreen!85!black] {$3$};

\node[paneltitle] at (12.10, 4.02) {(c) Bucket $B_2$ Routing};

\fill[NavyBlue!12, draw=NavyBlue!75, semithick] (12.35, 2.92) rectangle (14.85, 3.32);
\fill[RoyalBlue!18, draw=NavyBlue!75, semithick] (14.85, 2.92) rectangle (17.35, 3.32);
\node[font=\scriptsize\bfseries, text=NavyBlue!90!black] at (13.60, 3.12) {$S_2$};
\node[font=\scriptsize\bfseries, text=NavyBlue!90!black] at (16.10, 3.12) {$S_3$};

\draw[ForestGreen!85!black, very thick] (12.35, 2.84) -- (12.35, 3.40);
\node[font=\scriptsize\bfseries, text=ForestGreen!85!black, anchor=north] at (12.35, 2.82) {$p_8$};

\draw[BrickRed!90!black, very thick] (14.85, 2.84) -- (14.85, 3.40);
\node[font=\scriptsize\bfseries, text=BrickRed!90!black, anchor=north] at (14.85, 2.82) {$p_{11}$};

\draw[ForestGreen!85!black, very thick] (17.35, 2.84) -- (17.35, 3.40);
\node[font=\scriptsize\bfseries, text=ForestGreen!85!black, anchor=north] at (17.35, 2.82) {$p_{12}$};

\fill[Orange!90!black] (15.35, 3.12) circle (2.0pt);
\draw[-{Stealth[length=3.5pt, width=2.6pt]}, Orange!90!black, thick] (15.35, 3.60) -- (15.35, 3.22);
\node[font=\scriptsize\bfseries, text=Orange!90!black, anchor=south] at (15.35, 3.58) {$q$};

\draw[->, thick, black!65] (16.10, 2.78) -- (16.10, 2.32);

\draw[fill=black!3, draw=black!55, rounded corners=2pt, semithick] (12.35, 1.55) rectangle (17.35, 2.28);
\draw[black!35, thin] (14.85, 1.55) -- (14.85, 2.28);
\node[font=\tiny\bfseries, text=black!70, anchor=north] at (13.60, 2.24) {Left cover};
\node[font=\scriptsize\bfseries, text=NavyBlue!90!black, anchor=south] at (13.60, 1.60) {$S_2$};

\node[font=\tiny\bfseries, text=black!70, anchor=north] at (16.10, 2.24) {Internal split};
\node[font=\scriptsize\bfseries, text=BrickRed!85!black, anchor=south] at (16.10, 1.60) {$p_{11} \rightarrow S_3$};

\draw[->, thick, NavyBlue!85!black] (14.85, 1.50) -- (14.85, 1.08);
\node[font=\scriptsize, text=black!80, anchor=west] at (15.00, 1.30) {$\hat{r} = \lfloor S_3(q)\rfloor$};

\draw[black!45, semithick] (12.35, 0.62) -- (17.35, 0.62);
\fill[NavyBlue!16, draw=NavyBlue!80!black, semithick, rounded corners=1pt] (13.65, 0.48) rectangle (16.05, 0.76);
\fill[Orange!90!black] (14.85, 0.62) circle (1.9pt);
\node[font=\scriptsize\bfseries, text=NavyBlue!90!black, anchor=north] at (14.85, 0.44) {$[\hat{r}\!-\!\eps,\, \hat{r}\!+\!\eps]$};

\end{tikzpicture}
\caption{Piecewise-linear fitting, pivot promotion, and bucket routing. \textbf{(a)} Greedy PLA extends each segment until the first key ($p_5$) exits the $\pm\eps$ corridor, making $T_m$ a random stopping time. \textbf{(b)} On $N_0=19$ keys ($M_1=4$), Endpoint Promotion (EP) forwards segment starts with random strides $T_m$, while Rank-Equispaced Promotion (RE) steps by fixed stride $d_0 = \lfloor N_0/M_1\rfloor = 4$. \textbf{(c)} In RE bucket $B_2 = [p_8, p_{12})$ crossing segment start $p_{11}$, the directory stores left cover $S_2$ and split $p_{11}\to S_3$, routing $q$ to $\hat{r} = \lfloor S_3(q)\rfloor$.}
\label{fig:framework}
\end{figure*}
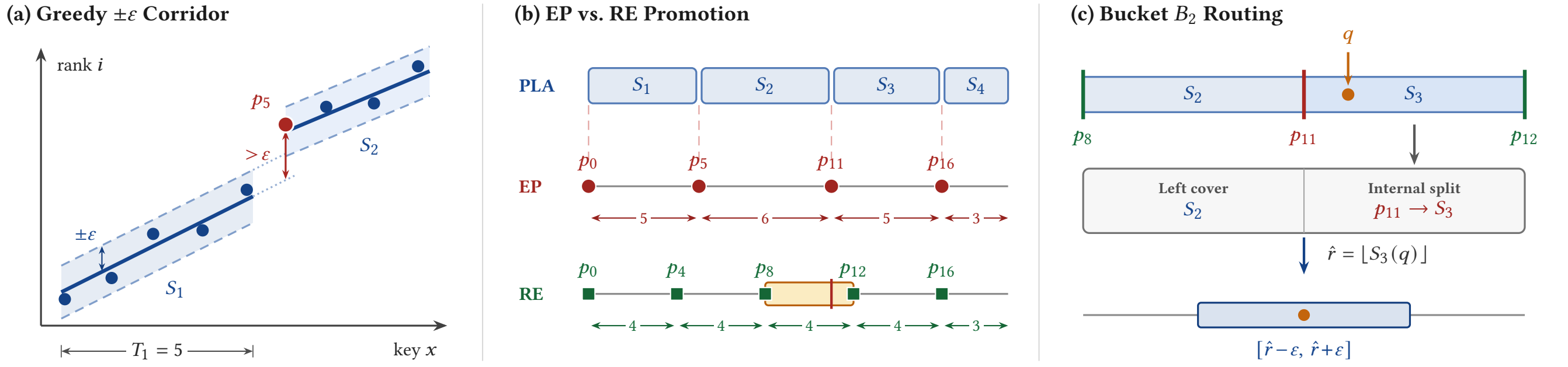

Learned indexes replace traditional B$^+$-tree page directories with lightweight models that predict where a query key lies in a sorted array~\cite{kraska2018case,bayer1972btree}.
Among them, piecewise-linear approximation (PLA) indexes have become a widely adopted paradigm across main-memory, persistent-memory, and domain-specific search engines~\cite{ferragina2020pgm,marcus2020sosd,jin2023plin,abrar2024plaindex} because they pair compact space with strict worst-case lookup bounds.
Every PLA learned index operates in two coupled stages (\Cref{fig:framework}).
First, a bottom-layer segmentation pass---originally developed for online time-series compression~\cite{orourke1981,elmeleegy2009swing,xie2014greedypla}---partitions the $n$ sorted keys into $M_1$ contiguous segments that approximate the empirical rank curve within error tolerance $\eps \ge 2$ (\Cref{fig:framework}(a)).
Second, an upper routing directory indexes those $M_1$ segments, allowing a query $q$ to locate its covering segment and finish with an $O(\log\eps)$ local search over at most $2\eps+1$ array positions.
Existing designs build this directory either by attaching an external structure---a B$^+$-tree in FITing-Tree~\cite{galakatos2019fittingtree}, a flat prefix table in RadixSpline~\cite{kipf2020radixspline}, or a multilevel histogram tree in PLEX~\cite{stoian2021plex}---or by recursively fitting new PLA layers over promoted segment representatives, as in the PGM-index~\cite{ferragina2020pgm}, PGM++~\cite{liu2025pgmpp}, and PLIN~\cite{jin2023plin}.

Classical interpolation search achieves $O(\log\log n)$ expected probes on uniform keys~\cite{perl1978interpolation} and extends to smooth positive densities via regularized interpolation~\cite{willard1985}.
More recently, learned-index theory has proved distribution-dependent $O(\log\log n)$ or $O(1)$ expected lookup for key-space piecewise-constant directories without worst-case per-segment error bounds~\cite{zeighami2023sublog,croquevielle2025constant}.
For error-bounded PLA indexes, however, prior stochastic analyses established only that a single optimal segment covers $\Theta(\eps^2)$ keys on average when key gaps fluctuate randomly~\cite{ferragina2020effective,qin2025pla}, stopping at one segment and coarse asymptotics.
Moving from a single segment to a full PLA index raises three fundamental questions: \emph{what exact constants govern single-layer coverage and when do fast one-pass fitters preserve $\Theta(\eps^2)$ scaling}, \emph{how does fanout evolve across recursive layers---does height stay logarithmic or collapse to $O(\log\log n)$}, and \emph{how do data skew and directory design interact across recursive, tree-based, and radix architectures?}

To answer these questions, we first establish a distribution-uniform Brownian limit at the bottom segmentation layer (\Cref{sec:single:brownian}).
Optimal greedy PLA covers $\kopt\eps^2/v_0 \approx 3.54\,\eps^2/v_0$ keys per segment on average under independent gaps ($v_0 = \sigma_0^2/\mu_0^2$) and $\kopt\eps^2$ under any smooth positive density $f$ ($v_0 = 1$).
We then track what happens when recursive PLA indexes such as the PGM-index~\cite{ferragina2020pgm} and PGM++~\cite{liu2025pgmpp} build upper layers by promoting one key (\emph{pivot}) from each segment and fitting new PLA segments to those pivots (\Cref{sec:standard:ep}).
Recent work on PGM++~\cite{liu2025pgmpp} (Lemma~3.2) claims that the standard PGM-index achieves $O(\log\log n)$ height, arguing via the Law of Large Numbers that summing $m = \Theta(\eps^2)$ gaps per segment reduces relative gap variance by $1/m$ at every layer.
We prove that under standard \emph{Endpoint Promotion (EP)}---which promotes the first key of each segment ($N_{\ell+1} = M_{\ell+1}$, \Cref{fig:framework}(b))---this $O(\log\log n)$ claim is incorrect.
The derivation in~\cite{liu2025pgmpp} treats the segment length $T_m$ as a deterministic count $m$, overlooking that each greedy segment terminates at the \emph{random stopping time} where the key path first exits the $\pm\eps$ corridor (\Cref{fig:framework}(a)).

Our Brownian limit shows that even as $\eps \to \infty$, the random stopping length $T_m$ retains a strictly positive relative variance $\aopt = \lim_{\eps\to\infty} \Var(T_m)/(\E[T_m])^2 \approx 0.17816$.
Instead of averaging down, these stopping-time fluctuations \emph{reset} the relative gap variance of every upper layer back to $\aopt$, erasing the initial key distribution within one to two layers.
Recursive fanout is therefore locked at a universal limit $\Ceps = (\kopt/\aopt)(\eps+\rhostar)^2 + \gamma + O(\eps^{-1}) \approx 19.861(\eps+0.257)^2 + 0.14$.
Consequently, standard EP never compounds its fanout across layers: its expected height is trapped at $\log n / \log \Ceps + O(1)$, and its expected lookup cost remains $\Theta(\log_\eps n \cdot \log\eps) = \Theta(\log n)$.
At practical tolerances such as $\eps = 8$, the variance-reset factor $\kopt/\aopt \approx 19.861$ inflates single-step fanout to $C_8 \approx 1{,}354$, so $\log_{1354} n$ spans only $2$ to $4$ levels up to $n = 10^9$.
This large base explains why empirical PGM heights in~\cite{liu2025pgmpp} appear nearly flat despite growing logarithmically in $n$.

Identifying stopping-time randomness as the bottleneck reveals how to unlock genuine doubly logarithmic scaling: \emph{decouple promoted pivots from segment boundaries} so that upper layers constructively realize the deterministic gap aggregation tacitly assumed in~\cite{liu2025pgmpp}.
We introduce \emph{Rank-Equispaced Promotion (RE)} (\Cref{fig:framework}(b)), which uses the segment count $M_{\ell+1}$ only to set a uniform integer rank stride $d_\ell = \floor{N_\ell / M_{\ell+1}}$ and promotes every $d_\ell$-th key in rank order ($N_{\ell+1} = \ceil{N_\ell / d_\ell}$).
Every full RE stride sums a \emph{deterministic} number $\Delta_\ell$ of original base gaps, so stopping-length fluctuations vanish identically and relative gap variance decays as $v_0 / \Delta_\ell$.
Each upper-layer segment now covers $d_\ell \asymp \eps^2 \Delta_\ell$ pivots, squaring cumulative rank coverage at every step ($\Delta_{\ell+1} = d_\ell \Delta_\ell \asymp \eps^2 \Delta_\ell^2$) and collapsing expected height to $\log_2\log_\eps n + O(1)$ on independent gaps (\Cref{sec:standard:gre}).

Analyzing RE on non-uniform key densities $f \in \Fclass_\delta$ uncovers a multi-layer curvature phase transition that is invisible at the bottom layer and masked by stopping-time noise under EP (\Cref{sec:standard:kre}).
Once RE's stride averaging suppresses stochastic gap noise below the macroscopic density curvature $\delta$ in the upper layers, remaining pivots contract via a square-root recurrence $N_{\ell+1} \asymp \sqrt{\delta N_\ell / \eps}$, producing a second doubly logarithmic phase of $\log_2^+(\log n / \log(\eps/\delta))$ layers.
Because RE pivots no longer align with segment starts (\Cref{fig:framework}(c)), we equip \textbf{RE-PGM} with a compact two-array bucket directory (\Cref{sec:routing:mech}).
A three-line counting argument shows that each segment boundary enters at most one bucket, bounding the expected routing overhead to \emph{strictly less than $1$ extra key comparison per layer} over stored keys.
RE-PGM thus achieves sub-logarithmic $O(\log\eps \cdot \log\log_\eps n)$ expected lookup in $\Theta(n/\eps^2)$ space and $O(n)$ build time.

Because bottom-layer segmentation and upper-layer routing decouple cleanly, our analysis extends beyond canonical PGM trees to online time-series compression and the broader ecosystem of PLA indexes and storage components (\Cref{sec:single:fitters,sec:routing:spectrum}).
At the bottom segmentation stage---which governs both index segment counts and streaming compression ratios---practical systems often replace optimal free-intercept PLA ($\kopt \approx 3.54$) with faster one-pass fitters.
We show that $\Theta(\eps^2/v_0)$ coverage is preserved whenever each segment anchors at a true data point (or restores interior cone margin~\cite{chen2026swing}): free-slope cone fitters (such as restarted Swing and anchored GreedyPLR variants~\cite{xie2014greedypla,chen2026moments,chen2026swing}) achieve the exact anchored constant $\kan = \frac{21}{\pi^2}\zeta(3) \approx 2.56$, and stricter prefix-chord fitters (\textsf{ShrinkingCone} in FITing-Tree~\cite{galakatos2019fittingtree} and \textsf{GreedySplineCorridor} in RadixSpline~\cite{kipf2020radixspline}) converge to a common Brownian chord constant $\kch \approx 1.09$.
In contrast, chaining segments through previous \emph{fitted} endpoints without resetting at a data point---as in the native Swing filter~\cite{elmeleegy2009swing,chen2026swing}---starves the initial slope-cone margin and degrades coverage to $\Theta(\eps\log\eps)$.

At the upper routing stage, combining these single-layer laws with our directory and height bounds yields unified expected space and lookup costs across both single-layer routing directories (FITing-Tree~\cite{galakatos2019fittingtree}, RadixSpline~\cite{kipf2020radixspline}, PLEX~\cite{stoian2021plex}) and recursive PLA hierarchies (PGM-index~\cite{ferragina2020pgm}, PGM++~\cite{liu2025pgmpp}, PLIN~\cite{jin2023plin}, RE-PGM), whether deployed standalone or inside LSM-trees and dynamic storage engines (\Cref{tab:family}).
This unified comparison makes the core architectural trade-off explicit.
Flat key-space directories such as RadixSpline~\cite{kipf2020radixspline} and equal-split piecewise-constant bins~\cite{croquevielle2025constant} partition \emph{key space}, achieving $O(\log\eps)$ or distribution-dependent $O(1)$ expected lookup on well-spread compact densities, yet their value bins suffer collisions under heavy key-space skew or wide support.
RE-PGM instead partitions \emph{rank space} to guarantee deterministic bucket occupancy, $\Theta(n/\eps^2)$ space, and $O(\log\eps \cdot \log\log_\eps n)$ lookup regardless of key-space skew.

\begin{table}[t]
\centering
\caption{Expected hierarchy height under optimal greedy PLA with tolerance $\eps$. Under EP, bounds hold uniformly for all $n \ge 2$ and $\eps \ge \eps_0$ with universal fanout $\Ceps = (\kopt/\aopt)\eps^2 + O(\eps) \approx 19.861\,\eps^2$. Under RE, bounds hold as $n, \eps \to \infty$ with $\log \eps = o(\log n)$, uniformly over the curvature-indexed density family $\Fclass_\delta$ ($\delta \in [0,1]$, $\log_2^+ x = \max(0, \log_2 x)$).}
\label{tab:matrix}
\footnotesize
\setlength{\tabcolsep}{4.5pt}
\begin{tabular}{@{}lll@{}}
\toprule
\textbf{Promotion Rule} & \textbf{Input Model} & \textbf{Expected Hierarchy Height $\E[H]$} \\
\midrule
Endpoint (EP) & Model~G (gap law $\nu_0$) & $\dfrac{\log n}{\log \Ceps} + O_{\nu_0}(1)$ \\[2.2ex]
Endpoint (EP) & Model~K (density $f$) & $\dfrac{\log n}{\log \Ceps} + O_f(1)$ \\[2.2ex]
Equispaced (RE) & Model~G (gap law $\nu_0$) & $\log_2\dfrac{\log n}{\log \eps} + O(1)$ \\[2.2ex]
Equispaced (RE) & Model~K ($f \in \Fclass_\delta$) & $\log_2\dfrac{\log n}{\log \eps} + \log_2^+\dfrac{\log n}{\log(\eps/\delta)} + O(1)$ \\
\bottomrule
\end{tabular}
\end{table}

\subsection{Contributions and Main Results}\label{sec:intro:contrib}\label{sec:intro:ideas}

We analyze error-bounded PLA indexing under two standard stochastic input models (\Cref{sec:prelim:data}): Model~G, where keys are partial sums of \iid\ positive gaps with mean $\mu_0$, variance $\sigma_0^2 > 0$, and finite third moment, and Model~K, where keys are sorted \iid\ draws from a positive bounded Lipschitz density $f$ on a compact interval.
Our main contributions are as follows (\Cref{tab:matrix,tab:family}):

\begin{itemize}[leftmargin=*,topsep=3pt plus 7pt,itemsep=3pt plus 7pt]
\item \textbf{Single-Layer Brownian Limits and Streaming Fitters (\Cref{sec:single}).}
We prove a distribution-uniform Donsker invariance principle tying discrete greedy PLA segment lengths to the Brownian affine-tube exit time $\tauopt$~\cite{chen2026moments} (\Cref{thm:single-F}), establishing exact $\Theta(\eps^2)$ fanout constants for optimal ($\kopt$), anchored ($\kan$), and prefix-chord ($\kch$) fitters (\Cref{sa:thm:comparison}) while proving that unanchored endpoint chaining in the native Swing filter degrades fanout to $\Theta(\eps\log\eps)$ (\Cref{thm:swing-scale}).

\item \textbf{Recursive EP Height Law and Universal Fanout (\Cref{sec:standard:ep}).}
We prove that stopping-time variance ($\aopt > 0$) resets upper-layer gap variance at every depth and locks recursive EP fanout at a distribution-free constant $\Ceps$ with an all-orders expansion in $\eps^{-1}$ (\Cref{thm:ceps}).
This establishes the uniform logarithmic height law $|\E[H_{\mathrm{ep}}] - \log n / \log \Ceps| \le K$ across all $n \ge 2$ under both Model~G and Model~K (\Cref{thm:height-ep}), refuting the $O(\log\log n)$ height claim in~\cite{liu2025pgmpp}.

\item \textbf{RE-PGM: Two-Phase Doubly Logarithmic Height and Directory Spectrum (\Cref{sec:standard:gre,sec:standard:kre,sec:routing:mech,sec:routing:spectrum}).}
We show that fixed-stride Rank-Equispaced Promotion (RE) eliminates stopping-time noise to achieve $\E[H_{\mathrm{re}}] = \log_2\log_\eps n + O(1)$ under Model~G (\Cref{thm:height-gre}) and the uniform two-phase height law $\log_2\log_\eps n + \log_2^+\frac{\log n}{\log(\eps/\delta)} + O(1)$ over the curvature-indexed density family $\Fclass_\delta$ (\Cref{thm:height-re}).
Coupled with a compact CSR bucket directory that adds under $1$ extra comparison per layer on average (\Cref{smq:lem:queried-bucket,thm:paper-re-index}), this yields unified cost bounds across single-layer and recursive directories (\Cref{tab:family}, Corollary~\ref{cor:body-directory-costs}).

\item \textbf{Empirical Validation and Lean~4 Formalization (\Cref{sec:prelim:lean,sec:numerics,app:lean-arch}).}
Over $10^8$-key benchmarks, our analytical constants predict empirical fanout within $0.3\%$ down to $\eps = 2$ and match exact integer heights (\Cref{sec:numerics}), while all eight core theorem packages are machine-checked without extra axioms in our 2,682-module Lean~4 library \texttt{PLAStochastic} (\Cref{app:lean-arch}).
\end{itemize}

\section{Preliminaries and Index Model}\label{sec:prelim}

We consider static predecessor search over $n$ sorted real keys $p_0 < p_1 < \dots < p_{n-1}$.
Given $q \in \R$, a predecessor query returns the largest index $i \in \{0, \dots, n-1\}$ with $p_i \le q$ (or reports $q < p_0$).

\subsection{Optimal Greedy PLA Segmentation}\label{sec:prelim:alg}

Let $p_0^{(\ell)} < p_1^{(\ell)} < \dots < p_{N_\ell - 1}^{(\ell)}$ denote the sorted keys at layer $\ell \ge 0$ of an index hierarchy, starting from the base array $p_i^{(0)} = p_i$ of size $N_0 = n$.
Fix an integer error tolerance $\eps \ge 2$.
A contiguous prefix of $T$ keys starting at index $s$ is \emph{$\eps$-feasible} if there exist $a, b \in \R$ predicting every local rank within $\pm\eps$:
\begin{equation}\label{eq:eps-feasible}
  \abs{(s + i) - \bigl(a p_{s+i}^{(\ell)} + b\bigr)} \le \eps
  \qquad\text{for all } 0 \le i < T.
\end{equation}
Condition~\eqref{eq:eps-feasible} requires all points $(p_{s+i}^{(\ell)}, s+i)$ to lie in a vertical strip of half-width $\eps$ around $y = ax + b$ (\Cref{fig:framework}(a)).

The optimal greedy segmentation algorithm (\textsf{ALG-OPT})~\cite{orourke1981,ferragina2020pgm} scans layer $\ell$ from left to right.
Starting at $s_1 = 0$, it finds the longest $\eps$-feasible prefix $T_1 \ge 2$, records the fitted linear model $S_1 = (a_1, b_1)$ covering indices $\{0, \dots, T_1 - 1\}$, and starts the next segment at the first rejected key $s_2 = T_1$.
Repeating this greedy scan until all $N_\ell$ keys are covered partitions layer $\ell$ into $M_{\ell+1}$ contiguous segments of lengths $T_1, \dots, T_{M_{\ell+1}}$ satisfying $\sum_{m=1}^{M_{\ell+1}} T_m = N_\ell$.
Maintaining the convex hull of feasible slopes via O'Rourke's online algorithm~\cite{orourke1981} constructs all $M_{\ell+1}$ segments in $O(N_\ell)$ time.

\subsection{Recursive Pivot Promotion: EP vs.\ RE}\label{sec:prelim:promotion}

When $M_{\ell+1} > 1$, a recursive PLA index promotes a subset of keys $p_0^{(\ell+1)} < \dots < p_{N_{\ell+1}-1}^{(\ell+1)}$ from layer $\ell$ to form layer $\ell+1$, repeating greedy segmentation until reaching a root $N_H = 1$ at height $H = \min\{\ell \ge 1 : N_\ell = 1\}$.
We compare two promotion rules (\Cref{fig:framework}(b)).

\textbf{Endpoint Promotion (EP).}
The standard PGM-index~\cite{ferragina2020pgm} promotes the first key of every segment:
\begin{equation}\label{eq:ep-def}
  N_{\ell+1} = M_{\ell+1},
  \qquad
  p_j^{(\ell+1)} = p_{s_{j+1}}^{(\ell)}
  \quad (0 \le j < M_{\ell+1}),
\end{equation}
where $s_{j+1} = \sum_{m=1}^{j} T_m$ is the starting index of segment $j+1$ in layer $\ell$.
Under EP, each promoted pivot is simultaneously a routing key and a segment start, so finding the predecessor among promoted pivots immediately identifies the segment covering any query $q \ge p_0$.
However, the rank spacing $T_j$ between consecutive promoted pivots $p_{j-1}^{(\ell+1)}$ and $p_j^{(\ell+1)}$ is the random stopping length of segment $j$.

\textbf{Rank-Equispaced Promotion (RE).}
To eliminate variable stopping lengths while preserving full key-range coverage, RE uses $M_{\ell+1}$ only to set a deterministic rank stride ($0 \le j < N_{\ell+1}$):
\begin{equation}\label{eq:re-def}
  d_\ell = \floor{N_\ell / M_{\ell+1}},
  \qquad
  N_{\ell+1} = \ceil{N_\ell / d_\ell},
  \qquad
  p_j^{(\ell+1)} = p_{j d_\ell}^{(\ell)}.
\end{equation}
By induction across layers, the $\ell$-th RE layer consists of original base keys spaced at cumulative rank stride $\Delta_\ell = \prod_{r=0}^{\ell-1} d_r$ (with $\Delta_0 = 1$), so that $p_j^{(\ell)} = p_{j\Delta_\ell}^{(0)}$ for all $0 \le j < N_\ell$.
The first $N_{\ell+1}-1$ buckets $[p_j^{(\ell+1)}, p_{j+1}^{(\ell+1)})$ each contain $d_\ell$ level-$\ell$ keys (spanning $\Delta_{\ell+1} = d_\ell \Delta_\ell$ original ranks), while the final tail bucket $[p_{N_{\ell+1}-1}^{(\ell+1)}, \infty)$ covers the remaining $N_\ell - (N_{\ell+1}-1)d_\ell \le d_\ell$ keys.
Because RE pivots do not coincide with segment starts, routing a query from bucket $j$ to its covering segment uses the lightweight bucket directory in \Cref{sec:routing}.

\subsection{Stochastic Input Models}\label{sec:prelim:data}

We analyze fanout, height, and query cost under two random input models separating local spacing noise from density variation.

\textbf{Model~G (Independent Positive Gaps).}
The base keys are partial sums $p_k = x_0 + \sum_{i=0}^{k-1} g_i$ ($0 \le k < n$) of \iid\ positive gap variables $g_0, g_1, \dots$ with mean $\mu_0 = \E[g_0] > 0$, variance $\sigma_0^2 = \Var(g_0) > 0$, relative variance $v_0 = \sigma_0^2 / \mu_0^2 > 0$, and finite third absolute central moment $\E[\abs{g_0 - \mu_0}^3] \le \kappa_0 < \infty$.
Because the empirical rank curve of Model~G has constant macroscopic slope $1/\mu_0$, Model~G isolates how local gap fluctuations interact with pivot promotion.

\textbf{Model~K (Bounded Lipschitz Key Density and $\Fclass_\delta$).}
The base keys $p_0 < p_1 < \dots < p_{n-1}$ are the order statistics of $n$ \iid\ samples drawn from a density $f$ on $[A, B]$ satisfying $0 < f_- \le f(x) \le f_+ < \infty$ and Lipschitz bound $\mathrm{Lip}(f) = \sup_{x \ne y}\abs{f(x) - f(y)}/\abs{x - y} \le L_f < \infty$ on $[A, B]$.
To study how non-uniform curvature affects RE height, we normalize $[A,B] = [0,1]$ and define for fixed $0 < f_- \le 1 \le f_+$, $L, V > 0$, and $0 < a < b < 1$ the \emph{curvature-indexed density family} $\Fclass_\delta$ ($\delta \in [0, 1]$):
\begin{equation}\label{eq:fclass-delta}
  \begin{aligned}
    \Fclass_\delta = \Bigl\{ f : [0,1] \to [f_-, f_+] \;\Big|\;
    &\textstyle\int_0^1 f = 1,\; \mathrm{Lip}(f) \le \delta L, \\
    &\abs{f(b) - f(a)} \ge \delta V \Bigr\}.
  \end{aligned}
\end{equation}
Here $\delta \in [0,1]$ controls the slope variation of $F(x) = \int_0^x f(u)\,du$: $\Fclass_0 = \{1\}$ is uniform on $[0,1]$, whereas $\delta > 0$ enforces both $\mathrm{Lip}(f) \le \delta L$ and non-flatness $\abs{f(b) - f(a)} \ge \delta V$ on $[a,b]$.

\begin{remark}[Model motivation, scaling boundaries, and parameter regimes]\label{rem:model-scope}
Models~G and~K capture complementary practical workloads without assuming uniform keys (uniformity corresponds only to the flat boundary case $\delta = 0$): Model~G models append-only timestamp, log, and arrival sequences with stationary local spacing fluctuations, whereas Model~K models continuous attribute values whose macroscopic density slope bends across $[A,B]$.
Their regularity hypotheses mark genuine scaling boundaries rather than proof artifacts.
Under heavy-tailed gaps with infinite variance, single-layer excursions no longer follow diffusive $\Theta(\eps^2)$ scaling, while densities that vanish ($f_- = 0$) or diverge ($f_+ = \infty$) break uniform local order-statistic spacing and bounded macroscopic curvature.
Outside these stochastic regimes, both EP and RE-PGM remain exact deterministic predecessor indexes with $O(n)$ construction time, worst-case height $O(\log_\eps n)$, and strictly fewer than $1$ expected extra bucket-routing comparison per layer over stored keys (\Cref{smq:lem:queried-bucket}).

Regarding parameter scaling, our EP height bounds hold for any fixed tolerance $\eps \ge \eps_0$ uniformly across all $n \ge 2$ (predicting empirical fanout within $0.3\%$ down to $\eps = 2$ in \Cref{sec:numerics}), whereas our RE bounds hold as $n, \eps \to \infty$ with $\log\eps = o(\log n)$.
Across the standard sub-polynomial range $\log\eps = o(\log n / \log\log n)$, RE-PGM achieves strictly sub-logarithmic $o(\log n)$ expected lookup, while for extreme tolerances $\log\eps = \Omega(\log n / \log\log n)$ the hierarchy already spans $O(\log\log n)$ levels and plain $O(\log n)$ binary search is asymptotically comparable.
\end{remark}

\subsection{Machine Verification in Lean~4}\label{sec:prelim:lean}

All core mathematical results of this paper---the Brownian affine lifetime moment recurrences and dispersion constants, the native Swing scaling laws, the unconditional EP additive height bounds under Models~G and~K, the all-orders asymptotic expansion of $\Ceps$, and the Model~G and uniform Model~K ($\Fclass_\delta$) RE height theorems---have been formally verified in Lean~4/Mathlib within our unified library \texttt{PLAStochastic}.
Each core theorem in \Cref{sec:single,sec:standard} cites its corresponding public Lean declaration, and \Cref{app:lean-arch} (\Cref{tab:lean-packages}) records the exact mapping between paper results, model hypotheses, and \texttt{PLAStochastic.Results} modules.
All formalized declarations depend solely on the three standard Lean~4 foundational axioms (\texttt{propext}, \texttt{Classical.choice}, and \texttt{Quot.sound}) with zero \texttt{sorry} placeholders or unproved analytical axioms.

\section{Single-Layer Fanout and the Fitter Spectrum}\label{sec:single}

Before analyzing multi-layer hierarchies, we must answer a basic question at a single layer: given a key sequence with relative gap variance $v_0 = \sigma_0^2 / \mu_0^2$ and error tolerance $\eps$, how many keys does a greedy PLA segment cover, and how much does that coverage fluctuate from segment to segment?

\subsection{The Brownian Functional Limit for Optimal Greedy PLA}\label{sec:single:brownian}

Consider fitting the first segment $T_\eps$ on $p_k = x_0 + \sum_{i=0}^{k-1} g_i$ under Model~G.
Why does an error budget $\eps$ always produce $\Theta(\eps^2/v_0)$ coverage?
If we normalize the mean gap to $\mu_0 = 1$, then $p_k = x_0 + k + \sum_{i=0}^{k-1}(g_i - 1)$ is a linear drift $k$ perturbed by a zero-mean random walk whose cumulative fluctuation over $m$ steps has standard deviation $\sqrt{m v_0}$.
Fitting a line $a p_k + b$ to the rank $k$ cancels the deterministic drift and shifts the origin, leaving only the centered random walk excursion.
For this excursion to stay inside a tube of half-width $\eps$, its typical fluctuation $\sqrt{m v_0}$ cannot exceed $\eps$, which forces the segment length to scale as $m \asymp \eps^2 / v_0$.

To turn this scaling heuristic into exact constants, we pass from the discrete centered walk to its continuous-time Gaussian scaling limit.
When index steps are rescaled by $m = t \eps^2 / v_0$ and vertical fluctuations are rescaled by $\eps$, Donsker's invariance principle~\cite{billingsley1999convergence} sends the linearly interpolated walk to a \emph{standard Brownian motion} $(B_t)_{t \ge 0}$---the continuous random path starting at $B_0 = 0$ with independent Gaussian increments $B_t - B_s \sim \mathcal{N}(0, t - s)$ for $0 \le s < t$.
Geometrically, $D_{\mathrm{opt}}(t)$ below is the smallest vertical half-width of any slanted strip enveloping $(s, B_s)_{0 \le s \le t}$, and $\tauopt$ is the continuous counterpart of greedy segmentation---the first-exit time at which no strip of half-width $1$ can contain the path:
\begin{equation}\label{eq:tauopt-def}
  D_{\mathrm{opt}}(t) = \inf_{a, b \in \R} \sup_{0 \le s \le t} \abs{B_s\!-\!a s\!-\!b},
  \;\;
  \tauopt = \inf\bigl\{ t \ge 0 : D_{\mathrm{opt}}(t) > 1 \bigr\}.
\end{equation}
Because $B_{c t}$ has the same distribution as $\sqrt{c}\, B_t$, the affine error satisfies $D_{\mathrm{opt}}(c t) \stackrel{d}{=} \sqrt{c}\, D_{\mathrm{opt}}(t)$, so exiting a corridor of half-width $\eps$ takes $\eps^2 \tauopt$ units of scaled time.

Upper layers of a recursive index apply \textsf{ALG-OPT} to promoted gap distributions that vary with both $\eps$ and layer depth $\ell$.
A pointwise limit for a single fixed gap law is therefore insufficient: convergence to $\tauopt$ must hold uniformly across an entire class of gap laws $\mathcal{G}(v_-, v_+, M_3)$ having positive relative variance $v(\nu) = \sigma^2/\mu^2 \in [v_-, v_+]$ and bounded normalized third moment $\E[\abs{g - \mu}^3]/\sigma^3 \le M_3$.

\begin{theorem}[Uniform Brownian Limit and Exact Lifetime Constants]\label{thm:single-F}
Fix $0 < v_- \le v_+ < \infty$ and $M_3 < \infty$.
As $\eps \to \infty$, the rescaled greedy segment length $v(\nu)\eps^{-2} T_\eps$ converges in distribution and in all moments of order $q < 3$ to $\tauopt$, uniformly over all gap laws $\nu \in \mathcal{G}(v_-, v_+, M_3)$:
\begin{equation}\label{eq:single-moments}
  \begin{aligned}
    \lim_{\eps \to \infty} \sup_{\nu \in \mathcal{G}(v_-, v_+, M_3)} \abs{\frac{v(\nu)}{\eps^2}\E_\nu[T_\eps] - \kopt} &= 0, \\
    \lim_{\eps \to \infty} \sup_{\nu \in \mathcal{G}(v_-, v_+, M_3)} \abs{\frac{\Var_\nu(T_\eps)}{(\E_\nu[T_\eps])^2} - \aopt} &= 0.
  \end{aligned}
\end{equation}
Moreover, the mean $\kopt = \E[\tauopt]$ and relative variance $\aopt = \Var(\tauopt)/\kopt^2$ are given in closed form by
\begin{equation}\label{eq:kopt-aopt}
  \begin{aligned}
    \kopt &= \frac{40}{\pi^2}\zeta(3) - \frac{4}{3} \approx 3.53842, \\
    \aopt &= \frac{\frac{2240}{\pi^4}\zeta(5) - \frac{224}{3\pi^2}\zeta(3)}{\kopt^2} - 1 \approx 0.17816.
  \end{aligned}
\end{equation}
\end{theorem}

The two constants in \eqref{eq:kopt-aopt} play complementary roles throughout the paper: $\kopt \approx 3.53842$ governs the expected fanout of a single layer, whereas $\aopt \approx 0.17816 > 0$ quantifies the irreducible randomness of greedy stopping times that drives multi-layer behavior in \Cref{sec:standard}.
Applying renewal theory to consecutive greedy segments immediately determines the total number of segments $M_1$ produced at the base layer:

\begin{corollary}[Base-Layer Segment Count]\label{cor:M-G}
Under Model~G with relative gap variance $v_0 = \sigma_0^2 / \mu_0^2$, as $n, \eps \to \infty$ with $\eps^2 = o(n)$, the base-layer segment count $M_1$ satisfies
\[
  \frac{M_1}{n v_0 / (\kopt \eps^2)} \xrightarrow{\Prob} 1
  \qquad\text{and}\qquad
  \frac{\E[M_1]}{n v_0 / (\kopt \eps^2)} \longrightarrow 1.
\]
Under Model~K, local uniformity of the density $f$ yields the same asymptotics with $v_0 = 1$ (since standardized order-statistic gaps behave locally like $\mathrm{Exp}(1)$ variables with unit relative variance).
\end{corollary}

The closed-form polynomial recurrences for $\E[\tauopt^q]$ in terms of odd zeta values $\zeta(3), \dots, \zeta(2q+1)$, together with the first four moments and $\aopt > 0$, are machine-checked in \leanid{PLAStochastic.Results.BrownianMoments} (\leanid{optimalUnitLifetime_moment_eq} and \leanid{optimalUnitLifetime_variance_eq_constant}).
The uniform discrete-to-continuous coupling is summarized in \Cref{app:brownian}.

\subsection{How Fitter Choices Affect Fanout: From Prefix Chords to Swing}\label{sec:single:fitters}

Many practical learned indexes replace O'Rourke's free-intercept convex hull scan with simpler streaming fitters that anchor each segment at a fixed starting point.
Do these faster fitters preserve the $\Theta(\eps^2)$ fanout law, or can a seemingly minor change in how a segment restarts degrade asymptotic space?

Prefix-chord fitters---namely \textsf{Shrinking\-Cone} in FITing-Tree~\cite{galakatos2019fittingtree} and \textsf{Greedy\-Spline\-Corridor} in RadixSpline~\cite{kipf2020radixspline}---anchor each new segment at the key-rank pair $(p_s, s)$ of its starting key and intersect feasible slope intervals online.
They accept a prefix of length $T$ whenever the chord joining $(p_s, s)$ to $(p_{s+T-1}, s+T-1)$ stays within $\pm\eps$ of all intermediate points.
A simple triangle-inequality comparison shows why anchoring at a true data point costs at most a constant factor: if a free line $(a,b)$ approximates a prefix within $\pm(\eps/2)$, shifting that line vertically by at most $\eps/2$ to pass through $(p_s, s)$ increases the maximum error on the prefix to at most $\eps/2 + \eps/2 = \eps$.

\begin{proposition}[Deterministic Two-Sided Fitter Bracket]\label{lem:envelope}\label{sa:thm:comparison}
On every deterministic key array $p_0 < \dots < p_{n-1}$ and for every tolerance $\eps \ge 2$, the greedy segment length $T^{\mathrm{ch}}(\eps)$ and total segment count $M^{\mathrm{ch}}(\eps)$ of any prefix-chord fitter (\textsf{ShrinkingCone} or \textsf{GreedySplineCorridor}) are bracketed between optimal greedy PLA at tolerances $\eps/2$ and $\eps$:
\begin{equation}\label{eq:chord-bracket}
  \begin{aligned}
    T^{\mathrm{opt}}(\floor{\eps/2}) &\le T^{\mathrm{ch}}(\eps) \le T^{\mathrm{opt}}(\eps), \\
    M^{\mathrm{opt}}(\eps) &\le M^{\mathrm{ch}}(\eps) \le M^{\mathrm{opt}}(\floor{\eps/2}).
  \end{aligned}
\end{equation}
Consequently, under Model~G or Model~K, prefix-chord fitters retain $\Theta(\eps^2)$ average fanout with limiting constant $\kch \in [\kopt/4, \kopt]$.
\end{proposition}

A strikingly different phenomenon occurs in the native Swing filter~\cite{elmeleegy2009swing,chen2026swing}, widely used for streaming time-series and continuous spline compression.
To produce a continuous piecewise-linear curve without storing separate intercepts, native Swing anchors segment $k+1$ not at the true data point $(t_e, y_e)$, but at the \emph{fitted endpoint} $(t_e, \hat{y}_e)$ of segment $k$, while carrying the rejected sample $t_e+1$ into segment $k+1$ as its first constraint.

Why does anchoring at the previous fitted value change the scaling order?
When segment $k$ terminates, its fitted line typically sits near the boundary of the $\pm E$ error corridor, leaving the new segment with a small initial margin $R_{k,E} = E - \abs{y_{t_e+1} - \hat{y}_{t_e}} = O(1)$ on one side of its slope cone.
On a Gaussian random walk $y_j = \sum_{i=0}^{j-1} X_i$ ($X_i \stackrel{\iid}{\sim} \mathcal{N}(0,1)$), a slope cone launched with an $O(1)$ margin has survival probability $\Theta(E / t)$ at intermediate times $t \in [E, E^2]$.
Integrating this heavy $\Theta(E / t)$ tail over $t \le E^2$ produces an expected segment length of order $\int_E^{E^2} (E/t)\,dt = \Theta(E \log E)$ rather than $\Theta(E^2)$.
Even more subtly, because the very first segment starts cleanly at $(0, y_0)$, its exit margin law $\rho$ puts positive mass on zero-drift overshoots that lack the $\log E$ tail, trapping the \emph{second} segment at $\Theta(E)$ before later segments $k \ge 3$ settle into the $\Theta(E \log E)$ regime:

\begin{theorem}[Native Swing Scaling Anomaly and Alternating Brackets~\cite{chen2026swing}]\label{thm:swing-scale}
Consider the native Swing streaming algorithm on a standard Gaussian random walk $y_j = \sum_{i=0}^{j-1} X_i$ ($X_i \stackrel{\iid}{\sim} \mathcal{N}(0,1)$, modeling detrended key-rank fluctuations) with error tolerance $E \to \infty$.
Let $L_{k,E}$ denote the number of samples covered by the $k$-th segment ($k \ge 1$). Then:
\begin{enumerate}[leftmargin=*, nosep]
  \item \emph{Three-regime scaling:} The first segment scales as $\E[L_{1,E}] \sim \kan E^2$ with $\kan = \frac{21}{\pi^2}\zeta(3) \approx 2.55767$, the second segment collapses to $\E[L_{2,E}] / E \to \kappa_2 > 0$ (with explicit finite-$E$ bounds $\frac{E}{24M} \le \E[L_{2,E}] \le C_{\mathrm{up}}E$ for all $E \ge 16$ and absolute constants $M, C_{\mathrm{up}} > 0$), and every subsequent segment $k \ge 3$ scales as
  \[
    \lim_{E \to \infty} \frac{\E[L_{k,E}]}{E \log E} = c_k > 0.
  \]
  \item \emph{Alternating convergence to steady state:} As $k \to \infty$, the coefficients $c_k$ bracket the unique stationary fanout constant $\gamma > 0$ from above on odd segments and from below on even segments, converging geometrically:
  \begin{equation}\label{eq:swing-brackets}
    \begin{aligned}
      c_4 \le c_6 \le \dots &\le \gamma \le \dots \le c_5 \le c_3, \\
      \abs{c_k - \gamma} &\le C_0 \left(\frac{3}{4}\right)^{k-3} \quad (k \ge 3),
    \end{aligned}
  \end{equation}
  for an explicit finite constant $C_0 > 0$ (\Cref{app:brownian}).
\end{enumerate}
\end{theorem}

\Cref{lem:envelope,thm:swing-scale} establish a clean design principle for learned-index builders: restarting each segment at an actual data key (even with a fast chord heuristic) preserves $\Theta(\eps^2)$ compression, whereas chaining segments through fitted endpoints to enforce continuity drops asymptotic fanout from $\Theta(\eps^2)$ to $\Theta(\eps \log \eps)$.
Both parts of \Cref{thm:swing-scale} are fully formalized in \leanid{PLAStochastic.Results.SwingScale} and \leanid{Results.SwingConstant} (declarations \leanid{swing_transient_coefficients}, \leanid{theorem_4_4_swing_gaussian}, and \leanid{swing_transient_brackets}).

\section{Recursive Height: Variance Reset vs.\ Two-Phase Scaling}\label{sec:standard}

We now turn to our central question: when a recursive PLA index builds successive layers using tolerance $\eps$, how does the promotion rule determine the height $H$ of the hierarchy?

\subsection{Endpoint Promotion: Why Variance Never Averages Out}\label{sec:standard:ep}

At first glance, promoting segment endpoints seems like it should rapidly smooth out the key distribution.
Each level-1 gap $G_j = p_j^{(1)} - p_{j-1}^{(1)} = \sum_{i=0}^{T_j - 1} g_{s_j + i}$ sums roughly $C = \E[T_j] \asymp \eps^2$ consecutive base gaps, and summing $C$ independent variables with relative variance $v_0$ normally reduces relative variance to $v_0 / C$.
If that $1/C$ variance reduction held under Endpoint Promotion, then layer $\ell$ would have relative variance $v_\ell \asymp \eps^{-2\ell}$, fanout would square at every step, and standard PGM-index height would be $O(\log\log n)$---precisely as claimed in the PGM++ analysis~\cite{liu2025pgmpp}.

Why does this argument fail completely for Endpoint Promotion?
The catch is that a greedy segment does not sum a fixed number of gaps.
It sums $T_j$ gaps up to the \emph{random stopping time} where the key path first exits the $\pm\eps$ corridor (\Cref{fig:framework}(a)).
Writing $\mu_0 = \E[g_i]$ and $C = \E[T_j]$, we can decompose the normalized promoted gap around $1$ into two asymptotically orthogonal components:
\begin{equation}\label{eq:ep-decomp}
  \frac{G_j}{\mu_0 C} - 1
  = \underbrace{\left(\frac{T_j}{C} - 1\right)}_{\text{stopping-length fluctuation}}
  + \underbrace{\frac{1}{\mu_0 C}\sum_{i=0}^{T_j - 1} (g_{s_j + i} - \mu_0)}_{\text{internal gap noise}}.
\end{equation}
By Wald's second-moment identity for the stopped martingale $\sum_{i=0}^{T_j-1}(g_{s_j+i} - \mu_0)$, the internal gap noise in the second term has variance $\frac{\sigma_0^2 \E[T_j]}{\mu_0^2 C^2} = \frac{v_0}{C} = O(\eps^{-2})$, which indeed vanishes as $\eps \to \infty$.
However, the first term measures the relative fluctuation of the stopping length $T_j$ itself.
By \Cref{thm:single-F}, $T_j / C$ converges in law and second moment to $\tauopt / \kopt$, whose variance is the strictly positive Brownian dispersion constant $\aopt = \Var(\tauopt)/\kopt^2 \approx 0.17816 > 0$.

\begin{figure}[t]
\centering
\begin{tikzpicture}
\begin{axis}[
    name=decomp,
    width=3.80cm,
    height=3.85cm,
    xmode=log,
    ymode=log,
    xmin=2, xmax=1000,
    ymin=0.0008, ymax=38.0,
    ytick={1e-3, 1e-2, 1e-1, 1e0},
    xlabel={Aggregation count $C$ or $\Delta$},
    ylabel={Relative variance $v_{\mathrm{next}}$},
    ylabel style={font=\scriptsize, yshift=-2pt},
    title={\textbf{(a) Variance Decomposition}},
    title style={font=\scriptsize, yshift=-3pt},
    label style={font=\scriptsize},
    tick label style={font=\scriptsize},
    grid=major,
    grid style={dashed, gray!25},
    legend image post style={xscale=0.45},
    legend style={font=\tiny, at={(0.97,0.97)}, anchor=north east, fill=white, fill opacity=0.92, draw=gray!40, inner sep=0.6pt, row sep=-1.6pt},
    legend cell align={left}
]
\addplot[BrickRed!90!black, very thick, domain=2:1000, samples=50] {0.17816 + 1.0/x};
\addlegendentry{EP ($\aopt+v_0/C$)}

\addplot[black!75, densely dashed, thick, domain=2:1000, samples=2] {0.17816};
\addlegendentry{Floor $\aopt$}

\addplot[ForestGreen!85!black, very thick, domain=2:1000, samples=50] {1.0/x};
\addlegendentry{RE ($v_0/\Delta$)}
\end{axis}

\begin{axis}[
    name=orbit,
    at={(decomp.east)},
    anchor=west,
    xshift=1.05cm,
    width=3.80cm,
    height=3.85cm,
    xmin=0, xmax=3,
    ymin=0.0, ymax=1.22,
    xtick={0,1,2,3},
    ytick={0, 0.5, 1.0},
    xticklabels={0,1,2,3},
    xlabel={Hierarchy layer $\ell$},
    ylabel={Empirical variance $v_\ell$},
    ylabel style={font=\scriptsize, yshift=-3pt},
    title={\textbf{(b) Multi-Layer Collapse}},
    title style={font=\scriptsize, yshift=-3pt},
    label style={font=\scriptsize},
    tick label style={font=\scriptsize},
    grid=major,
    grid style={dashed, gray!25},
    legend image post style={xscale=0.45},
    legend style={font=\tiny, at={(0.97,0.97)}, anchor=north east, fill=white, fill opacity=0.92, draw=gray!40, inner sep=0.6pt, row sep=-1.8pt},
    legend cell align={left}
]
\addplot[NavyBlue!90!black, thick, mark=*, mark size=1.3pt] coordinates {
    (0, 1.000) (1, 0.188) (2, 0.176) (3, 0.177)
};
\addlegendentry{Exp ($\eps=4$)}

\addplot[RoyalBlue!85!black, thick, mark=square*, mark size=1.2pt] coordinates {
    (0, 1.000) (1, 0.178) (2, 0.178) (3, 0.179)
};
\addlegendentry{Exp ($\eps=16$)}

\addplot[Orange!90!black, thick, mark=triangle*, mark size=1.4pt] coordinates {
    (0, 0.333) (1, 0.191) (2, 0.178) (3, 0.178)
};
\addlegendentry{Unif ($\eps=4$)}

\addplot[BrickRed!85!black, thick, mark=diamond*, mark size=1.4pt] coordinates {
    (0, 0.333) (1, 0.179) (2, 0.179) (3, 0.178)
};
\addlegendentry{Unif ($\eps=16$)}

\addplot[ForestGreen!85!black, very thick, densely dashdotted, mark=pentagon*, mark size=1.4pt] coordinates {
    (0, 1.000) (1, 0.0010) (2, 0.00001) (3, 0.00001)
};
\addlegendentry{RE ($v_0/\Delta_\ell$)}

\addplot[black!80, dashed, semithick, domain=0:3] {0.17816};
\node[font=\tiny\bfseries, text=black!85, anchor=south] at (axis cs:2.0, 0.20) {$\aopt \approx 0.178$};
\end{axis}
\end{tikzpicture}
\caption{Why endpoint promotion (EP) resets variance while rank-equispaced promotion (RE) eliminates it. \textbf{(a)} By decomposition~\eqref{eq:ep-decomp}, internal gap noise decays as $v_0/C$, but EP retains the stopping-length variance floor $\aopt = \Var(\tauopt)/\kopt^2 \approx 0.178$, whereas RE steps by a fixed stride $\Delta$ and follows $v_0/\Delta$. \textbf{(b)} Empirical gap variance across layers ($\mathrm{Exp}(1)$ and $\mathrm{Unif}[0,2]$ at $\eps=4, 16$): all four EP trajectories collapse within two layers to $\aopt \approx 0.178$, while RE drives relative variance to zero after one promotion.}
\label{fig:recursion}
\end{figure}
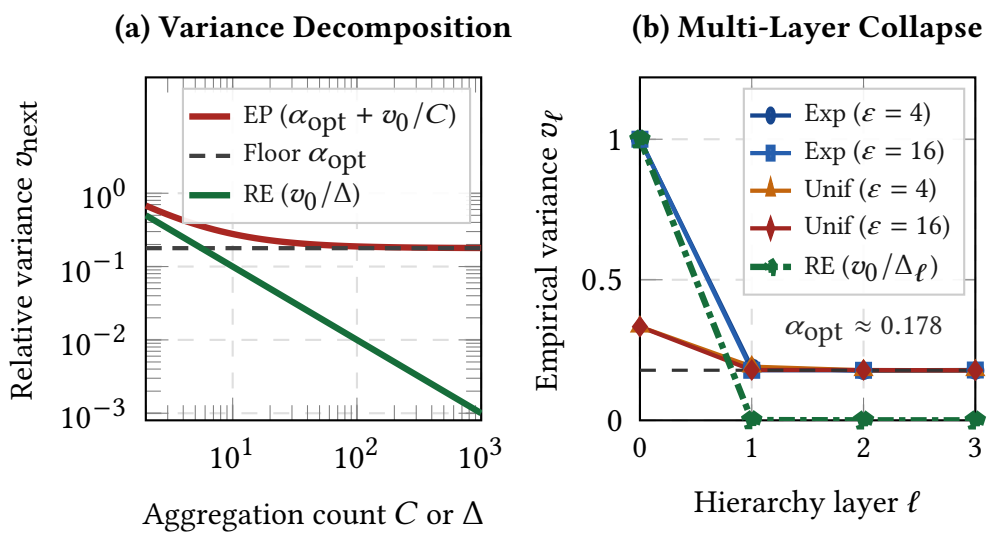

As shown in \Cref{fig:recursion}(a--b), no matter what relative variance $v_0$ the base keys start with, the random stopping length $T_j$ injects fresh relative variance $\aopt \approx 0.17816$ into the promoted gaps at the very first layer:

\begin{proposition}[Single-Step Variance Reset under EP]\label{prop:variance-reset}
Uniformly over all base gap laws $\nu \in \mathcal{G}(v_-, v_+, M_3)$, the promoted gap $G_1 = \sum_{i=0}^{T_\eps - 1} g_i$ under Endpoint Promotion satisfies
\begin{equation}\label{eq:var-reset}
  \lim_{\eps \to \infty} \sup_{\nu \in \mathcal{G}(v_-, v_+, M_3)}
  \abs{\frac{\Var_\nu(G_1)}{(\E_\nu[G_1])^2} - \aopt} = 0.
\end{equation}
Substituting this reset variance $v_{\mathrm{next}} \to \aopt$ into the single-layer fanout formula $\kopt \eps^2 / v_{\mathrm{next}}$ of \Cref{thm:single-F} drives the asymptotic fanout of every upper layer $\ell \ge 1$ to the universal constant
\begin{equation}\label{eq:cinf-def}
  \Cinf(\eps) = \frac{\kopt}{\aopt}\eps^2 \approx 19.86098\,\eps^2.
\end{equation}
\end{proposition}

What happens at finite $\eps \ge \eps_0$ across arbitrarily many layers?
Let $\Phi_\eps(\nu)$ denote the exact distribution of the normalized promoted gap $G_1 / \E_\nu[G_1]$ when \textsf{ALG-OPT} is applied to \iid\ $\nu$-gaps, and let $C_\eps(\nu) = \E_\nu[T_\eps]$ be the expected single-layer fanout.
Iterating the layer operator $\Phi_\eps$ from the exponential reference law $\mathrm{Exp}(1)$ defines the exact limiting fanout constant
\begin{equation}\label{eq:ceps-def}
  \log \Ceps = \lim_{j \to \infty} \log C_\eps\bigl(\Phi_\eps^j(\mathrm{Exp}(1))\bigr).
\end{equation}
Because $\Phi_\eps$ contracts rapidly across layers (\Cref{fig:recursion}(b)), the actual layer-mean fanouts $C_\ell(\eps)$ converge geometrically to $\Ceps$, locking the hierarchy height to $\log n / \log \Ceps$ within $O(1)$ layers for \emph{all} $n \ge 2$:

\begin{theorem}[Unconditional Logarithmic Height under EP]\label{thm:height-ep}\label{thm:height-G}\label{thm:height-K}
Under either Model~G (with fixed gap law $\nu_0$) or Model~K (with fixed bounded Lipschitz density $f$ on $[A,B]$), there exist constants $\eps_0, K > 0$ depending only on the input law such that for all $\eps \ge \eps_0$ and all $n \ge 2$,
\begin{equation}\label{eq:ep-height-bound}
  \abs{\E[H_{\mathrm{ep}}] - \frac{\log n}{\log \Ceps}} \le K,
\end{equation}
where $\Ceps > 1$ is the universal compression constant defined in \eqref{eq:ceps-def}. No asymptotic growth condition relating $n$ and $\eps$ is required.
\end{theorem}

Moreover, $\Ceps / \eps^2$ admits a unique all-orders asymptotic expansion in powers of $\eps^{-1}$, whose first-order correction captures the discrete ladder-height overshoot of the key walk when a segment exits the $\pm\eps$ corridor:

\begin{theorem}[Asymptotic Expansion of Universal EP Fanout $\Ceps$]\label{thm:ceps}
There exists a unique sequence of real coefficients $a_0, a_1, a_2, \dots$ with leading constant $a_0 = \kopt / \aopt \approx 19.86098$ such that for every integer $M \ge 0$, there exist constants $\eps_M, K_M > 0$ satisfying
\begin{equation}\label{eq:ceps-expansion}
  \abs{\frac{\Ceps}{\eps^2} - \sum_{j=0}^{M} a_j \eps^{-j}} \le K_M \eps^{-(M+1)}
  \qquad\text{for all } \eps \ge \eps_M.
\end{equation}
\end{theorem}

Analytically, the first-order coefficient is given by $a_1 = 2 a_0 \rhostar \in (10.2123631452682,\, 10.2123631452684)$, where $\rhostar \approx 0.25710$ is the ladder-height overshoot constant of the normalized Brownian reference law $\Lambda = \mathrm{Law}(\tauopt / \kopt)$.
Consequently, the shifted-square expansion $\Ceps = a_0(\eps + \rhostar)^2 + \gamma + O(\eps^{-1}) = a_0\eps^2 + a_1\eps + a_2 + O(\eps^{-1})$ (with $a_0\rhostar^2 \approx 1.31278$, empirical residual $\gamma \approx 0.14$, and $a_2 = a_0\rhostar^2 + \gamma \approx 1.45$) matches empirical stationary fanout within $0.3\%$ across all $\eps \in [2, 64]$ (\Cref{sec:numerics}).

Turning the single-layer variance reset \eqref{eq:ep-decomp} into the unconditional all-$n$ height bound \eqref{eq:ep-height-bound} requires proving that finite-$\eps$ deviations from $\Ceps$ do not accumulate over arbitrarily deep hierarchies---specifically, establishing the cross-layer summability bound $\sum_{j \ge 1} \abs{\log C_j(\eps) - \log \Ceps} \le K'$ under Model~G and coupling local Poisson windows to the exponential reference orbit under Model~K.
We summarize these multi-layer coupling arguments and the explicit integral evaluation of $a_1 = 2a_0\rhostar$ in \Cref{app:ep-proofs}.
The complete proofs of \Cref{thm:height-ep,thm:ceps} (including existence, universality, and all-orders expansion of $\Ceps$ with $a_0 = \kopt/\aopt$) are machine-checked in \texttt{PLAStochastic} (declarations \leanid{PGM.H_G_EP_additive_law}, \leanid{PGM.H_K_EP_additive_law}, \leanid{PGM.sum_abs_log_segMean_sub_le}, and \leanid{PGM.gepC_asymptoticExpansion}).

Our variance-reset results pinpoint where the $O(\log\log n)$ proof in~\cite{liu2025pgmpp} breaks and why canonical PGM nonetheless appears flat in practice.
In~\cite{liu2025pgmpp}, Lemma~3.1 conditions on prior-layer segment coverages $C_0 = c_0, \dots, C_{i-1} = c_{i-1}$ and treats each level-$i$ gap as a sum of a deterministic count $m = c_0 \cdots c_{i-1}$ of raw base gaps.
Substituting $\Var(\sum_{r=1}^m g_r) / (m\mu_0)^2 = v_0 / m$ then yields the claimed $\E[C_i] \ge (\eps^2/v_0)^{2^i}$ fanout bound (Lemma~3.2), $O(\log_2\log_G n)$ static height (Theorem~3.3 and cost-model Eq.~(18e)), and $O(\log n \log\log n)$ dynamic lookup for DPGM-Index.
Footnote~2 of~\cite{liu2025pgmpp} notes that the derivation assumes equal coverage across segments within a level and suggests invoking concentration bounds, yet \Cref{thm:single-F,prop:variance-reset} show why concentration cannot repair this step: $T_j / \E[T_j]$ does not concentrate around $1$ as $\eps \to \infty$, but converges to the non-degenerate limit $\tauopt / \kopt$ with relative variance $\aopt \approx 0.17816 > 0$.

At the same time, \Cref{thm:ceps} explains the motivating benchmark in Table~4 of~\cite{liu2025pgmpp}, where canonical PGM height at $\eps = 8$ on uniform keys rose only from $2$ to $4$ across $n = 10^3$ to $10^9$ (versus $4$ to $11$ for a B$^+$-tree of fanout $B = 8$), with over $99.9\%$ of segments in the bottom layer.
Variance reset amplifies upper-layer fanout by $1/\aopt \approx 5.61$ over the base layer, driving stationary EP fanout at $\eps = 8$ to $C_8 \approx 19.861(8 + 0.257)^2 + 0.14 \approx 1{,}354$---nearly $170\times$ larger than $B = 8$.
Consequently, $\log_{1354}(10^9) \approx 2.87$ adds only one level every three orders of magnitude in $n$, while all upper levels combined hold fewer than $1/C_8 < 0.1\%$ of the segments.

PGM++~\cite{liu2025pgmpp} pairs this hierarchy with three effective systems optimizations: hybrid linear/branchless-\texttt{CMOV} local search, skipping sparse top layers below a fixed segment threshold, and separate internal and leaf tolerances $(\eps_i, \eps_\ell)$.
These techniques substantially cut CPU branch misses and constant-factor latency, and branchless hybrid search applies orthogonally to RE-PGM as well.
Asymptotically, however, for any fixed $(\eps_i, \eps_\ell)$ and fixed bypass threshold, top-layer bypassing removes $O(1)$ levels while the internal EP hierarchy still contracts by $C_{\eps_i}$ per level in $\log n / \log C_{\eps_i} + O(1)$ layers.

\subsection{Rank-Equispaced Promotion: Deterministic Strides and Doubly Logarithmic Height}\label{sec:standard:gre}

Decomposition~\eqref{eq:ep-decomp} immediately reveals how to unlock compounding variance reduction across layers: eliminate the stopping-length fluctuation $(T_j/C - 1)$ by choosing promoted pivots at a \emph{fixed rank stride}.
Under Rank-Equispaced Promotion~\eqref{eq:re-def}, layer $\ell$ consists of base keys $p_{j\Delta_\ell}^{(0)}$ spaced at cumulative rank stride $\Delta_\ell = \prod_{r=0}^{\ell-1} d_r$.

Why does a fixed rank stride collapse hierarchy height from $\Theta(\log_\eps n)$ to $\log_2\log_\eps n + O(1)$?
On any fixed grid of rank stride $\Delta_\ell$, each level-$\ell$ gap $G_j^{(\ell)} = \sum_{i=0}^{\Delta_\ell - 1} g_{(j-1)\Delta_\ell + i}$ is the sum of a \emph{deterministic} number $\Delta_\ell$ of independent base gaps.
The aggregation count is non-random, so the first term of \eqref{eq:ep-decomp} is identically zero.
Only the internal gap noise remains, and the relative gap variance at layer $\ell$ drops inversely with the cumulative rank stride:
\begin{equation}\label{eq:re-var-decay}
  v_\ell = \frac{\Var(G_j^{(\ell)})}{(\E[G_j^{(\ell)}])^2} = \frac{v_0}{\Delta_\ell}.
\end{equation}
Substituting $v_\ell = v_0 / \Delta_\ell$ into the single-layer fanout formula of \Cref{thm:single-F} shows that each linear segment at layer $\ell$ covers $d_\ell \asymp \frac{\kopt \eps^2}{v_\ell} = A \Delta_\ell$ pivots, where $A = \kopt \eps^2 / v_0$.
Since each level-$\ell$ pivot already represents $\Delta_\ell$ base ranks, the cumulative rank stride squares at every layer:
\begin{equation}\label{eq:re-squaring}
  \Delta_{\ell+1} = d_\ell \Delta_\ell \asymp A \Delta_\ell^2
  \quad\Longrightarrow\quad
  A \Delta_\ell \asymp (A \Delta_0)^{2^\ell} = \left(\frac{\kopt \eps^2}{v_0}\right)^{2^\ell}.
\end{equation}
Each promotion step doubles $\log(A\Delta_\ell)$, so $\Delta_H$ reaches $n$ in $\log_2(\log n / \log(A)) \approx \log_2\log_\eps n$ layers:

\begin{theorem}[Doubly Logarithmic RE Height under Model~G]\label{thm:height-gre}
Under Model~G, as $n, \eps \to \infty$ with $\log \eps = o(\log n)$, the expected height of Rank-Equispaced Promotion satisfies
\begin{equation}\label{eq:gre-height}
  \E[H_{\mathrm{re}}] = \log_2\left(\frac{\log n}{\log \eps}\right) + O(1),
\end{equation}
and more precisely $\E[H_{\mathrm{re}}] - \log_2(\log n / \log \eps) \in [-1, 0] + o(1)$, or equivalently $\E[H_{\mathrm{re}}] - y_n \in [0, 1] + o(1)$ for $y_n = \log_2\bigl(\frac{\log n}{\log(\kopt \eps^2 / v_0)} + 1\bigr)$ with $H_{\mathrm{re}} = \ceil{y_n}$ outside an $o(1)$ rounding window.
\end{theorem}

Making the squaring recursion \eqref{eq:re-squaring} rigorous involves one subtle probabilistic obstacle: in the actual algorithm, the grid stride $\Delta_\ell$ is not fixed in advance, but determined dynamically from the data via $d_{r} = \floor{N_r / M_{r+1}}$.
Conditioning naively on $\Delta_\ell$ would destroy the independence of the base gaps.
Our proof overcomes this by establishing a uniform concentration event simultaneously across all reachable grid strides $\Delta \in \{1, \dots, n\}$.
The complete argument (along with exact outside-window integer formulas) is summarized in \Cref{app:re-proofs} and fully machine-checked in \texttt{PLAStochastic.Results.GRE} (declarations \leanid{PGM.G_RE_exact}, \leanid{PGM.G_RE_exact_integral}, and \leanid{PGM.G_RE_exact_limit}).

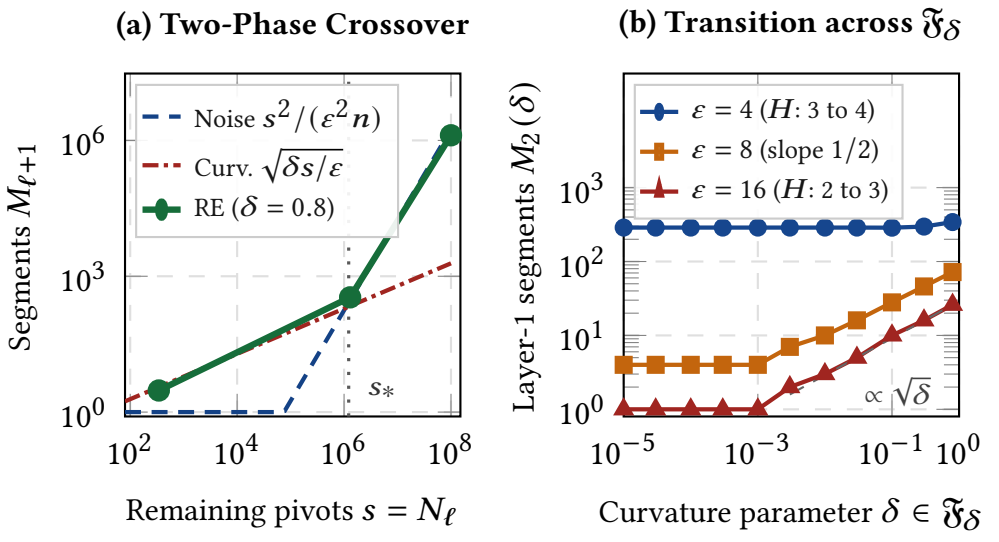
\begin{figure}[t]
\centering
\begin{tikzpicture}
\begin{axis}[
    name=crossover,
    width=3.80cm,
    height=3.85cm,
    xmode=log,
    ymode=log,
    xmin=80, xmax=1.5e8,
    ymin=0.8, ymax=3e7,
    xlabel={Remaining pivots $s = N_\ell$},
    ylabel={Segments $M_{\ell+1}$},
    ylabel style={font=\scriptsize, yshift=-2pt},
    title={\textbf{(a) Two-Phase Crossover}},
    title style={font=\scriptsize, yshift=-3pt},
    label style={font=\scriptsize},
    tick label style={font=\scriptsize},
    xtick={1e2, 1e4, 1e6, 1e8},
    ytick={1e0, 1e3, 1e6},
    grid=major,
    grid style={dashed, gray!25},
    legend image post style={xscale=0.42},
    legend style={font=\tiny, at={(0.03,0.97)}, anchor=north west, fill=white, fill opacity=0.92, draw=gray!40, inner sep=0.6pt, row sep=-1.6pt},
    legend cell align={left}
]
\addplot[NavyBlue!85!black, thick, densely dashed, domain=80:1e8, samples=50] {max(1.0, (x^2) / (56.615 * 1e8))};
\addlegendentry{Noise $s^2/(\eps^2 n)$}

\addplot[BrickRed!85!black, thick, densely dashdotted, domain=80:1e8, samples=50] {0.193 * sqrt(x)};
\addlegendentry{Curv.\ $\sqrt{\delta s/\eps}$}

\addplot[ForestGreen!85!black, very thick, mark=*, mark size=1.5pt] coordinates {
    (100000000, 1297888)
    (1298702, 342)
    (343, 3)
};
\addlegendentry{RE ($\delta = 0.8$)}

\draw[black!65, dotted, semithick] (axis cs:1.22e6, 0.8) -- (axis cs:1.22e6, 3e7);
\node[font=\tiny\bfseries, text=black!80, anchor=west] at (axis cs:1.35e6, 3.0) {$s_*$};
\end{axis}

\begin{axis}[
    name=deltascan,
    at={(crossover.east)},
    anchor=west,
    xshift=1.08cm,
    width=3.80cm,
    height=3.85cm,
    xmode=log,
    ymode=log,
    xmin=1e-5, xmax=1.0,
    ymin=0.8, ymax=3.5e4,
    xlabel={Curvature parameter $\delta \in \Fclass_\delta$},
    ylabel={Layer-1 segments $M_2(\delta)$},
    ylabel style={font=\scriptsize, yshift=-3pt},
    title={\textbf{(b) Transition across $\Fclass_\delta$}},
    title style={font=\scriptsize, yshift=-3pt},
    label style={font=\scriptsize},
    tick label style={font=\scriptsize},
    xtick={1e-5, 1e-3, 1e-1, 1},
    ytick={1, 10, 100, 1000},
    grid=major,
    grid style={dashed, gray!25},
    legend image post style={xscale=0.42},
    legend style={font=\tiny, at={(0.03,0.97)}, anchor=north west, fill=white, fill opacity=0.92, draw=gray!40, inner sep=0.6pt, row sep=-1.6pt},
    legend cell align={left}
]
\addplot[NavyBlue!90!black, thick, mark=*, mark size=1.3pt] coordinates {
    (1e-5, 288) (3e-5, 288) (1e-4, 288) (3e-4, 288) (1e-3, 288)
    (3e-3, 288) (1e-2, 288) (3e-2, 288) (1e-1, 287) (3e-1, 297) (8e-1, 342)
};
\addlegendentry{$\eps = 4$ ($H$: 3 to 4)}

\addplot[Orange!90!black, thick, mark=square*, mark size=1.3pt] coordinates {
    (1e-5, 4) (3e-5, 4) (1e-4, 4) (3e-4, 4) (1e-3, 4)
    (3e-3, 7) (1e-2, 10) (3e-2, 16) (1e-1, 28) (3e-1, 46) (8e-1, 72)
};
\addlegendentry{$\eps = 8$ (slope $1/2$)}

\addplot[BrickRed!85!black, thick, mark=triangle*, mark size=1.5pt] coordinates {
    (1e-5, 1) (3e-5, 1) (1e-4, 1) (3e-4, 1) (1e-3, 1)
    (3e-3, 2) (1e-2, 3) (3e-2, 5) (1e-1, 10) (3e-1, 16) (8e-1, 26)
};
\addlegendentry{$\eps = 16$ ($H$: 2 to 3)}

\addplot[black!60, densely dashed, thin, domain=3e-3:0.8] {29.0 * sqrt(x)};
\node[font=\tiny, text=black!75, anchor=north west] at (axis cs:0.022, 4.8) {$\propto\sqrt{\delta}$};
\end{axis}
\end{tikzpicture}
\caption{Two-phase noise-to-curvature transition across the uniform density class $\Fclass_\delta$ ($n=10^8$). \textbf{(a)} At layer size $s=N_\ell$, stochastic order-statistic noise requires $M_{\mathrm{noise}}(s)\propto s^2/(\eps^2 n)$ segments (slope $2$), whereas second-order Taylor curvature of $sF(x)$ requires $M_{\mathrm{curv}}(s)\propto\sqrt{\delta s/\eps}$ segments (slope $1/2$). Once RE drives $s$ below the crossover $s_*$, the trajectory switches from squaring its stride to square-root compression. \textbf{(b)} Measured layer-1 segment count $M_2(\delta)$ across $\delta \in [10^{-5}, 0.8]$: a flat noise plateau for small $\delta$ transitions into $\sqrt{\delta/\eps}$ curvature growth, activating the second $\log_2^+\frac{\log n}{\log(\eps/\delta)}$ height phase.}
\label{fig:curvature}\label{fig:routing}
\end{figure}

\subsection{Macroscopic Curvature and the Uniform Two-Phase Law across \texorpdfstring{$\Fclass_\delta$}{F\_delta}}\label{sec:standard:kre}

When keys are drawn from a non-uniform density $f$ on $[0,1]$ (Model~K), can RE keep squaring its rank stride all the way to the root?
Suppose layer $\ell$ retains $s = N_\ell \approx n / \Delta_\ell$ equispaced rank pivots $p_j^{(\ell)} \approx F^{-1}(j/s)$.
As illustrated in \Cref{fig:curvature}(a), fitting these $s$ pivots within rank error $\pm\eps$ faces two competing lower bounds on the required segment count $M_{\ell+1}$:
\begin{enumerate}[leftmargin=*, nosep]
  \item \emph{Stochastic order-statistic noise:} Each gap between consecutive level-$\ell$ pivots aggregates $\Delta_\ell \approx n/s$ exponential order-statistic spacings, giving relative variance $v_\ell \approx 1/\Delta_\ell = s/n$. By \Cref{cor:M-G}, local stochastic fluctuations alone require
  \begin{equation}\label{eq:m-noise}
    M_{\mathrm{noise}}(s) \asymp \frac{s \cdot v_\ell}{\kopt \eps^2} = \frac{s^2}{\kopt \eps^2 n}.
  \end{equation}
  \item \emph{Macroscopic density curvature:} Even if all stochastic noise were removed so that $p_j^{(\ell)}$ lay exactly on the smooth curve $y(x) = s F(x)$, approximating $y(x)$ by a straight line on an interval of width $h$ within $\pm\eps$ incurs a second-order Taylor sagitta error of $\frac{1}{16}\abs{y''(x)}h^2 = \frac{1}{16}s\abs{f'(x)}h^2 \le \eps$. When $f \in \Fclass_\delta$ has slope variation of scale $\delta$, each segment can span width at most $h \asymp \sqrt{\eps / (\delta s)}$, requiring at least
  \begin{equation}\label{eq:m-curv}
    M_{\mathrm{curv}}(s) \asymp \sqrt{\frac{\delta s}{\eps}}.
  \end{equation}
\end{enumerate}

Comparing \eqref{eq:m-noise} and \eqref{eq:m-curv} reveals a clean two-phase trajectory (\Cref{fig:curvature}(a)):
\begin{itemize}[leftmargin=*, nosep]
  \item \textbf{Phase~I (Noise-Dominated Squaring, $s > s_*$):} In early layers where $s$ is large, $M_{\mathrm{noise}}(s) \gg M_{\mathrm{curv}}(s)$. Stochastic noise dominates, and each layer squares the stride: $s_{\ell+1} \asymp \frac{s_\ell^2}{\kopt \eps^2 n}$. This phase reduces $s_\ell$ down to the crossover scale $s_* \asymp (\delta \eps^3 n^2)^{1/3}$ in $\log_2\frac{\log n}{\log\eps} + O(1)$ layers.
  \item \textbf{Phase~II (Curvature-Dominated Square-Root Compression, $s < s_*$):} Once $s_\ell$ drops below $s_*$, noise has been averaged away and macroscopic curvature takes over ($M_{\mathrm{curv}}(s) \gg M_{\mathrm{noise}}(s)$). Now each layer follows the deterministic square-root recurrence
  \begin{equation}\label{eq:curv-recurrence}
    s_{\ell+1} \asymp M_{\mathrm{curv}}(s_\ell) \asymp \sqrt{\frac{\delta s_\ell}{\eps}}
    \quad\Longrightarrow\quad
    \log\left(\frac{\eps s_{\ell+1}}{\delta}\right) \approx \frac{1}{2}\log\left(\frac{\eps s_\ell}{\delta}\right).
  \end{equation}
  The positive log-ratio $\log(\eps s_\ell / \delta)$ halves at every step from $s_0 \le n$ until the penultimate layer reaches $s_\ell \le \eps / \delta$ (which compresses to a single root segment at the next step), adding $\log_2^+\frac{\log n}{\log(\eps/\delta)} + O(1)$ layers.
\end{itemize}

Crucially, this two-phase mechanism holds \emph{uniformly} across the entire curvature-indexed density family $\Fclass_\delta$ ($\delta \in [0,1]$) defined in \eqref{eq:fclass-delta}:

\begin{theorem}[Uniform Two-Phase RE Height Law across $\Fclass_\delta$]\label{thm:height-kre}\label{thm:height-re}\label{lem:landau-curv}
Fix density-class parameters $0 < f_- \le 1 \le f_+$, $L, V > 0$, and $0 < a < b < 1$ defining $\Fclass_\delta$ on $[0,1]$.
There exists a constant $K > 0$ depending only on the class parameters $(f_-, f_+, L, V, a, b)$ such that as $n, \eps \to \infty$ with $\log \eps = o(\log n)$, uniformly over all $\delta \in [0, 1]$ and all densities $f \in \Fclass_\delta$,
\begin{equation}\label{eq:kre-uniform}
  \begin{aligned}
    &\abs{\E[H_{\mathrm{re}}] - \Phi(n, \eps, \delta)} \le K, \\
    &\Phi(n, \eps, \delta) = \log_2\left(\frac{\log n}{\log \eps}\right) + \log_2^+\left(\frac{\log n}{\log(\eps/\delta)}\right),
  \end{aligned}
\end{equation}
where $\log_2^+ x = \max(0, \log_2 x)$ (taken to be $0$ at $\delta = 0$).
In particular:
\begin{enumerate}[leftmargin=*, nosep]
  \item When $\delta = 0$ (uniform density $f \equiv 1$) or $\delta \le \eps / n$, the second term vanishes and $\E[H_{\mathrm{re}}] = \log_2\frac{\log n}{\log\eps} + O(1)$.
  \item For any fixed non-constant Lipschitz density ($\delta > 0$ fixed as $n, \eps \to \infty$ with $\log\eps = o(\log n)$), both phases contribute equally and
  \[
    \E[H_{\mathrm{re}}] = 2\log_2\left(\frac{\log n}{\log \eps}\right) + O_f(1).
  \]
\end{enumerate}
\end{theorem}

\Cref{thm:height-kre} explains a striking contrast between the two promotion rules on skewed data.
Under EP, local stopping-time noise $\aopt \approx 0.178$ is so large at every layer that macroscopic curvature never alters the leading denominator $\log\Ceps$ in $\E[H_{\mathrm{ep}}] = \log n / \log \Ceps + O_f(1)$, and empirical height stays flat across $\delta$ (\Cref{fig:numerics}(c)).
Under RE, rapid noise cancellation strips away stochastic fluctuations within a few layers, exposing the underlying geometry of $F(x)$ and shifting the asymptotic coefficient from $1$ to $2$ as $\delta$ increases (\Cref{fig:curvature}(b)).
Both the uniform two-phase theorem over $\Fclass_\delta$ and the fixed-density exact height theorem are fully formalized in \texttt{PLAStochastic.Results.KRE} (declarations \leanid{PGM.K_RE_uniform}, \leanid{PGM.K_RE_uniform_integral}, \leanid{PGM.K_RE_fixed_density}, and \leanid{PGM.K_RE_exact_integral}, with full proof details in \Cref{app:re-proofs}).

\section{Directory Routing, Cost Spectrum, and Numerical Evaluation}\label{sec:routing}\label{sec:experiments}

Reducing hierarchy height from $\Theta(\log_\eps n)$ to $O(\log\log_\eps n)$ is only useful if queries can step from one RE layer to the next without hidden search overhead.
In this section, we first analyze the exact space and comparison cost of RE bucket routing and place six PLA index architectures on a common cost spectrum (\Cref{sec:routing:mech,sec:routing:spectrum}).
We then evaluate our theoretical predictions against empirical segmentations across finite $\eps$, $n$, and curvature $\delta$ (\Cref{sec:numerics}).

\subsection{Boundary-Crossing Bucket Routing}\label{sec:routing:mech}

Under Endpoint Promotion, each level-$(\ell+1)$ pivot is the starting key of a level-$\ell$ segment, so finding the predecessor of a query key $q$ among level-$(\ell+1)$ pivots directly identifies the unique level-$\ell$ segment covering $q$.
Under Rank-Equispaced Promotion, pivots $p_j^{(\ell+1)} = p_{j d_\ell}^{(\ell)}$ are placed at fixed rank intervals $d_\ell = \floor{N_\ell / M_{\ell+1}}$, so the $j$-th RE bucket $B_{\ell, j} = [p_j^{(\ell+1)}, p_{j+1}^{(\ell+1)})$ can straddle $b_{\ell, j} \ge 0$ internal segment starts of layer $\ell$ (\Cref{fig:framework}(c)).

To resolve which segment covers $q \in B_{\ell, j}$, each layer stores a compact Compressed Sparse Row (CSR) bucket directory consisting of two flat arrays:
\begin{enumerate}[leftmargin=*, nosep]
  \item A header array of length $N_{\ell+1}$, where entry $j$ stores the index $L_{\ell, j}$ of the level-$\ell$ segment covering the left endpoint $p_j^{(\ell+1)}$, together with an offset into a boundary array of length at most $M_{\ell+1}-1$.
  \item A contiguous slice of $b_{\ell, j}$ internal segment starts $p_{s_m}^{(\ell)} \in (p_j^{(\ell+1)}, p_{j+1}^{(\ell+1)})$ that fall strictly inside bucket $B_{\ell, j}$.
\end{enumerate}

When a query reaches bucket $B_{\ell, j}$, if $b_{\ell, j} = 0$, no segment boundary lies inside the bucket and segment $L_{\ell, j}$ is selected with zero key comparisons.
If $b_{\ell, j} \ge 1$, the query scans or binary-searches the $b_{\ell, j}$ internal boundary keys to select the covering segment $S_m = (a_m, b_m)$, evaluates its affine model $\hat{r} = \floor{a_m q + b_m}$ clamped to $[s_m, s_{m+1}-1] \cap [j d_\ell, (j+1)d_\ell - 1]$, and performs binary search over the local correction window $[\hat{r} - \eps - 1, \hat{r} + \eps + 1]$.

Could an adversary or a burst of short segments cause $b_{\ell, j}$ to blow up and erase RE's height advantage?
A three-line deterministic counting argument rules this out when queries are averaged uniformly over the $n$ stored keys:

\begin{lemma}[Amortized Boundary Comparisons per Layer]\label{smq:lem:queried-bucket}
On any deterministic key array $p_0 < \dots < p_{n-1}$, let $j_\ell(p_i)$ denote the level-$\ell$ RE bucket containing stored key $p_i$.
Then at every layer $\ell \ge 0$, the total number of internal boundary keys across all buckets satisfies $\sum_{j=0}^{N_{\ell+1}-1} b_{\ell, j} \le M_{\ell+1} - 1$, and the average number of internal boundaries in the queried bucket is strictly less than $1$:
\begin{equation}\label{eq:bucket-bound}
  \frac{1}{n}\sum_{i=0}^{n-1} b_{\ell, j_\ell(p_i)}
  \le \frac{d_\ell \Delta_\ell (M_{\ell+1}\!-\!1)}{n}
  \le \frac{(N_\ell\!-\!1) \Delta_\ell}{n} < 1.
\end{equation}
\end{lemma}
\begin{proof}
Each of the $M_{\ell+1}-1$ internal segment starts $p_{s_m}^{(\ell)}$ ($m \ge 2$) belongs to the interior $(p_j^{(\ell+1)}, p_{j+1}^{(\ell+1)})$ of at most one bucket $B_{\ell, j}$, which yields $\sum_j b_{\ell, j} \le M_{\ell+1}-1$.
Every bucket $B_{\ell, j}$ covers at most $d_\ell$ level-$\ell$ keys and hence at most $d_\ell \Delta_\ell$ original stored keys $p_i$.
Weighting bucket $j$ by its maximum occupancy $d_\ell \Delta_\ell / n$ and using $d_\ell (M_{\ell+1} - 1) \le N_\ell - d_\ell \le N_\ell - 1$ (since $d_\ell \ge 1$) together with $(N_\ell - 1)\Delta_\ell < n$ gives \eqref{eq:bucket-bound}.
\end{proof}

Combining \Cref{smq:lem:queried-bucket} with our height bounds (\Cref{thm:height-gre,thm:height-kre}) and the single-layer segment count $M_1 = \Theta(n/\eps^2)$ (\Cref{cor:M-G}) yields the end-to-end complexity of the RE-PGM index:

\begin{theorem}[RE-PGM Space, Construction, and Query Complexity]\label{thm:paper-re-index}
On any sorted array of $n$ keys with tolerance $\eps \ge 2$, the complete-coverage RE-PGM index returns the exact predecessor for every query $q \in \R$, occupies $O(\sum_{\ell \ge 0} M_{\ell+1})$ words of directory space, and builds in $O(n)$ deterministic time.
Under Model~G or Model~K ($f \in \Fclass_\delta$), as $n, \eps \to \infty$ with $\log\eps = o(\log n)$, the index uses $\Theta(n/\eps^2)$ expected words of auxiliary space and answers predecessor queries in
\begin{equation}\label{eq:re-query-cost}
  O\bigl(\log\eps \cdot \log\log_\eps n\bigr)
  \quad\text{expected operations and key comparisons}
\end{equation}
when averaged uniformly over the $n$ stored keys.
\end{theorem}

Across the standard sub-polynomial tolerance range $\log\eps = o(\log n / \log\log n)$, the lookup bound~\eqref{eq:re-query-cost} is strictly sub-logarithmic ($o(\log n)$) while preserving $\Theta(n/\eps^2)$ auxiliary space and worst-case local search window $2\eps + 3$ at every layer.

\begin{table*}[t]
\centering
\caption{Comparison of six piecewise-linear index architectures under Model~G and Model~K ($\eps \ge 2$, $\log\eps = o(\log n)$, query cost averaged uniformly over the $n$ stored keys, with PGM++ shown at fixed $\eps_i = \eps_\ell = \eps$ and fixed scan/bypass threshold). Here $\Ceps = (\kopt/\aopt)\eps^2 + O(\eps) \approx 19.861\,\eps^2$, and $\Phi(n,\eps,\delta) = \log_2\frac{\log n}{\log\eps} + \log_2^+\frac{\log n}{\log(\eps/\delta)}$.}
\label{tab:family}
\footnotesize
\setlength{\tabcolsep}{4.8pt}
\begin{tabular}{@{}llllll@{}}
\toprule
\textbf{Index Architecture} & \textbf{PLA Fitter} & \textbf{Directory / Promotion} & \textbf{Auxiliary Space} & \textbf{Directory Levels} & \textbf{Expected Lookup Cost} \\
\midrule
PGM-index (EP)~\cite{ferragina2020pgm} & Optimal (\textsf{ALG-OPT}) & Recursive Endpoint (EP) & $\Theta(n / \eps^2)$ & $\dfrac{\log n}{\log \Ceps} + O(1)$ & $\Theta(\log_\eps n \cdot \log\eps) = \Theta(\log n)$ \\[1.6ex]
PGM++~\cite{liu2025pgmpp} & Optimal (\textsf{ALG-OPT}) & EP + fixed-threshold hybrid/bypass & $\Theta(n / \eps^2)$ & $\dfrac{\log n}{\log \Ceps} + O(1)$ & $O(\log_\eps n \cdot \log\eps) = O(\log n)$ \\[1.6ex]
\textbf{RE-PGM (Ours)} & Optimal (\textsf{ALG-OPT}) & Recursive Equispaced (RE) & $\Theta(n / \eps^2)$ & $\Phi(n,\eps,\delta) + O(1)$ & $O\bigl(\log\eps \cdot \log\log_\eps n\bigr)$ \\[1.2ex]
\midrule
FITing-Tree~\cite{galakatos2019fittingtree} & \textsf{ShrinkingCone} & B$^+$-tree (fanout $B$) & $\Theta(n / \eps^2)$ & $\Theta\bigl(\log_B(n/\eps^2)\bigr)$ & $O\bigl(\log(n/\eps^2) + \log\eps\bigr)$ \\[1.2ex]
RadixSpline~\cite{kipf2020radixspline} & \textsf{GreedySplineCorridor} & Flat $r$-bit radix table & $\Theta(n/\eps^2 + 2^r)$ & $1$ (flat table) & $O(\log\eps)$ at $2^r = \Theta(n/\eps^2)$ \\[1.2ex]
PLEX / PLIN~\cite{stoian2021plex,jin2023plin} & Spline / Optimal PLA & Multi-level radix / B$^+$-tree & $\Theta(n / \eps^2)$ & $O(1)$ radix / $\Theta(\log_B(n/\eps^2))$ & $O(\log\eps)$ to $O(\log(n/\eps^2)+\log\eps)$ \\
\bottomrule
\end{tabular}
\end{table*}

\subsection{The PLA Directory Cost Spectrum}\label{sec:routing:spectrum}

Because all static PLA indexes share the same bottom-layer segmentation problem ($M_1 = \Theta(n/\eps^2)$ segments of average length $\Theta(\eps^2)$), our single-layer brackets (\Cref{cor:M-G,lem:envelope}) and recursive height laws (\Cref{thm:height-ep,thm:height-re}) place the major PLA index families on a common analytical footing (\Cref{tab:family}):

\begin{corollary}[Cross-Architecture Cost Spectrum]\label{cor:body-directory-costs}
Under Model~G or Model~K as $n, \eps \to \infty$ with $\log\eps = o(\log n)$, the auxiliary space and stored-key expected lookup cost of the static PLA directories in \Cref{tab:family} satisfy:
\begin{enumerate}[leftmargin=*, nosep]
  \item \emph{Recursive PLA directories (PGM-index, PGM++, RE-PGM):} Stopping-time variance $\aopt > 0$ prevents fanout compounding under EP, so standard EP requires $\log n / \log \Ceps + O(1)$ layers and $\Theta(\log_\eps n \cdot \log\eps) = \Theta(\log n)$ expected comparisons, and PGM++~\cite{liu2025pgmpp} with fixed tolerances $(\eps_i, \eps_\ell)$ and fixed scan/bypass thresholds traverses $\log n / \log C_{\eps_i} + O(1)$ internal EP layers in $O(\log n)$ expected operations. In contrast, RE-PGM achieves $\Phi(n,\eps,\delta) + O(1)$ layers and $O(\log\eps \cdot \log\log_\eps n)$ expected comparisons in $\Theta(n/\eps^2)$ space.
  \item \emph{Tree-indexed PLA (FITing-Tree~\cite{galakatos2019fittingtree}, static PLIN directory~\cite{jin2023plin}):} Indexing $M_1 = \Theta(n/\eps^2)$ static segments with a B$^+$-tree of page fanout $B$ uses $\Theta(n/\eps^2)$ words and $O(\log(n/\eps^2) + \log\eps)$ comparisons.
  \item \emph{Radix-indexed PLA (RadixSpline~\cite{kipf2020radixspline}, PLEX~\cite{stoian2021plex}):} Under Model~K with bounded density ratio $f_+/f_- = O(1)$, a flat radix table of $2^r = \Theta(n/\eps^2)$ entries contains $O(1)$ spline points per radix bucket on average, achieving $O(\log\eps)$ expected lookup in $\Theta(n/\eps^2)$ space.
\end{enumerate}
\end{corollary}

\Cref{tab:family} highlights a fundamental trade-off between partitioning \emph{key space} and partitioning \emph{rank space}.
When keys come from a smooth compact density with bounded ratio $f_+/f_- = O(1)$, RadixSpline's flat $2^r$-entry prefix table~\cite{kipf2020radixspline} places $O(1)$ spline knots per bin at $2^r = \Theta(n/\eps^2)$ and achieves $O(\log\eps)$ expected lookup without traversing a hierarchy.
Under heavy key-space skew ($f_+/f_- \gg 1$ or clustered prefixes), however, uniform value bins overflow and require multi-level fallbacks as in PLEX~\cite{stoian2021plex}.
RE-PGM avoids this skew sensitivity by partitioning \emph{rank space} ($p_j^{(\ell+1)} = p_{j d_\ell}^{(\ell)}$) over $\pm\eps$ PLA segments: its bucket occupancy is deterministically at most $d_\ell$ regardless of key-space skew or support width, trading an $O(\log\log_\eps n)$ hierarchy factor for skew-invariant $\Theta(n/\eps^2)$ space and strict $O(\log\eps)$ worst-case local search per layer.

The same key-space geometry governs piecewise-constant RMI/PCA directories~\cite{zeighami2023sublog} and the Equal-Split Piecewise Constant (ESPC) index of Croquevielle et al.~\cite{croquevielle2025constant}.
ESPC divides $[p_0, p_{n-1}]$ into $K$ equal-width value bins, predicts each bin's midpoint rank, and finishes with unbounded exponential search in expected time $O\bigl(\log((B-A)\rho_f \, n/K) + 1\bigr)$ on $[A,B]$, where $\rho_f = \int_A^B f(x)^2\,dx$.
Rather than achieving a distribution-free constant, ESPC's $K = n$ bound $O(\log((B-A)\rho_f) + 1)$ grows with R\'enyi density concentration, requires superlinear $O(n\log n)$ to $O(n\sqrt{n\log n})$ space on unbounded support~\cite{croquevielle2025constant}, and provides no worst-case $\pm\eps$ search bound.
At matched PLA space $K = \Theta(M_1)$, ESPC yields $O\bigl(\log\eps + \log((B-A)\rho_f)\bigr)$ expected lookup under Model~K and $O\bigl(\log(\eps^2/v_0 + 2 + v_0)\bigr)$ under Model~G (by Lorden's renewal bound~\cite{lorden1970}), remaining sensitive to key-space concentration $(B-A)\rho_f$ whereas RE-PGM guarantees $\Theta(n/\eps^2)$ space and $O(\log\eps \cdot \log\log_\eps n)$ expected lookup without unbounded value-bin overflow.

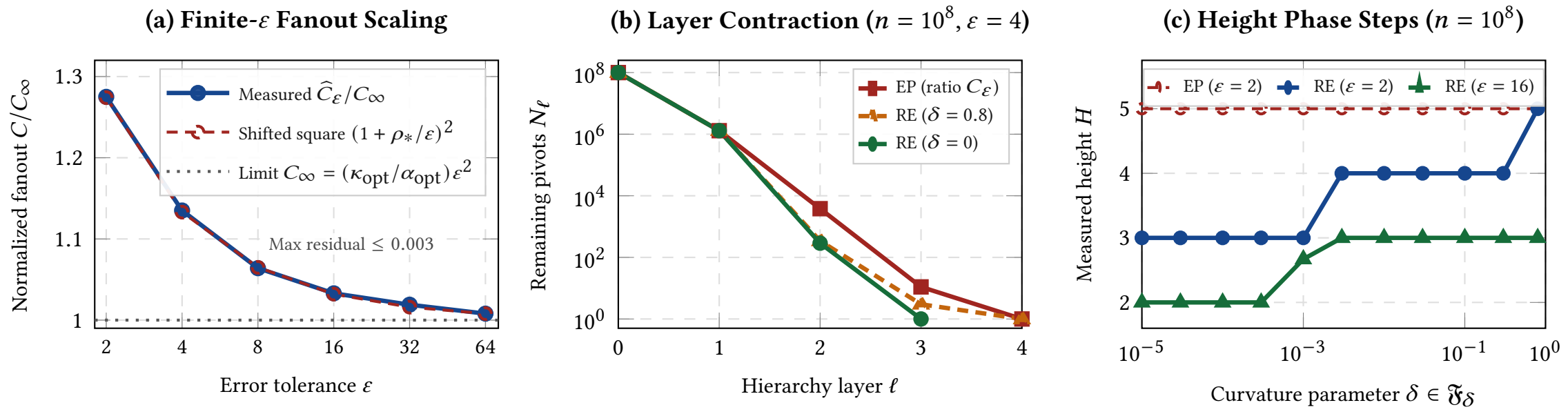
\begin{figure*}[t]
\centering
\begin{tikzpicture}
\begin{axis}[
    name=panelA,
    width=5.45cm,
    height=4.15cm,
    xmode=log,
    log basis x=2,
    xmin=1.8, xmax=72,
    ymin=0.99, ymax=1.32,
    xtick={2,4,8,16,32,64},
    xticklabels={2,4,8,16,32,64},
    xlabel={Error tolerance $\eps$},
    ylabel={Normalized fanout $C / C_\infty$},
    title={\textbf{(a) Finite-$\eps$ Fanout Scaling}},
    title style={font=\footnotesize, yshift=-2pt},
    label style={font=\scriptsize},
    tick label style={font=\scriptsize},
    grid=major,
    grid style={dashed, gray!25},
    legend style={font=\tiny, at={(0.97,0.97)}, anchor=north east, fill=white, fill opacity=0.92, draw=gray!40, inner sep=1.2pt},
    legend cell align={left}
]
\addplot[NavyBlue!90!black, very thick, mark=*, mark size=1.6pt] coordinates {
    (2, 1.275) (4, 1.135) (8, 1.064) (16, 1.033) (32, 1.019) (64, 1.008)
};
\addlegendentry{Measured $\widehat{C}_\eps / C_\infty$}

\addplot[BrickRed!85!black, thick, densely dashed, mark=o, mark size=1.6pt] coordinates {
    (2, 1.274) (4, 1.133) (8, 1.065) (16, 1.032) (32, 1.016) (64, 1.008)
};
\addlegendentry{Shifted square $(1+\rhostar/\eps)^2$}

\addplot[black!65, dotted, thick, domain=1.8:72] {1.0};
\addlegendentry{Limit $C_\infty = (\kopt/\aopt)\eps^2$}

\node[font=\tiny, text=black!80, fill=white, fill opacity=0.9, inner sep=1pt, anchor=south west]
  at (axis cs:8.5, 1.08) {Max residual $\le 0.003$};
\end{axis}

\begin{axis}[
    name=panelB,
    at={(panelA.east)},
    anchor=west,
    xshift=1.15cm,
    width=5.45cm,
    height=4.15cm,
    ymode=log,
    xmin=0, xmax=4,
    ymin=0.5, ymax=2.5e8,
    xtick={0,1,2,3,4},
    ytick={1, 1e2, 1e4, 1e6, 1e8},
    xlabel={Hierarchy layer $\ell$},
    ylabel={Remaining pivots $N_\ell$},
    title={\textbf{(b) Layer Contraction ($n=10^8, \eps=4$)}},
    title style={font=\footnotesize, yshift=-2pt},
    label style={font=\scriptsize},
    tick label style={font=\scriptsize},
    grid=major,
    grid style={dashed, gray!25},
    legend image post style={xscale=0.45},
    legend style={font=\tiny, at={(0.97,0.97)}, anchor=north east, fill=white, fill opacity=0.92, draw=gray!40, inner sep=0.8pt, row sep=-1.5pt},
    legend cell align={left}
]
\addplot[BrickRed!85!black, very thick, mark=square*, mark size=1.5pt] coordinates {
    (0, 100000000) (1, 1297888) (2, 3810) (3, 11) (4, 1)
};
\addlegendentry{EP (ratio $\Ceps$)}

\addplot[Orange!90!black, very thick, densely dashed, mark=triangle*, mark size=1.7pt] coordinates {
    (0, 100000000) (1, 1298702) (2, 343) (3, 3) (4, 1)
};
\addlegendentry{RE ($\delta = 0.8$)}

\addplot[ForestGreen!85!black, very thick, mark=*, mark size=1.5pt] coordinates {
    (0, 100000000) (1, 1298702) (2, 289) (3, 1)
};
\addlegendentry{RE ($\delta = 0$)}
\end{axis}

\begin{axis}[
    name=panelC,
    at={(panelB.east)},
    anchor=west,
    xshift=1.15cm,
    width=5.45cm,
    height=4.15cm,
    xmode=log,
    xmin=1e-5, xmax=1.0,
    ymin=1.6, ymax=5.75,
    xtick={1e-5, 1e-3, 1e-1, 1},
    ytick={2,3,4,5},
    xlabel={Curvature parameter $\delta \in \Fclass_\delta$},
    ylabel={Measured height $H$},
    title={\textbf{(c) Height Phase Steps ($n=10^8$)}},
    title style={font=\footnotesize, yshift=-2pt},
    label style={font=\scriptsize},
    tick label style={font=\scriptsize},
    grid=major,
    grid style={dashed, gray!25},
    legend columns=3,
    legend image post style={xscale=0.35},
    legend style={font=\tiny, at={(0.5,0.97)}, anchor=north, fill=white, fill opacity=0.92, draw=gray!40, inner sep=0.6pt, column sep=1.5pt},
    legend cell align={left}
]
\addplot[BrickRed!85!black, thick, densely dashed, mark=o, mark size=1.4pt] coordinates {
    (1e-5, 5) (3e-5, 5) (1e-4, 5) (3e-4, 5) (1e-3, 5)
    (3e-3, 5) (1e-2, 5) (3e-2, 5) (1e-1, 5) (3e-1, 5) (8e-1, 5)
};
\addlegendentry{EP ($\eps = 2$)}

\addplot[NavyBlue!90!black, very thick, mark=*, mark size=1.4pt] coordinates {
    (1e-5, 3) (3e-5, 3) (1e-4, 3) (3e-4, 3) (1e-3, 3)
    (3e-3, 4) (1e-2, 4) (3e-2, 4) (1e-1, 4) (3e-1, 4) (8e-1, 5)
};
\addlegendentry{RE ($\eps = 2$)}

\addplot[ForestGreen!85!black, very thick, mark=triangle*, mark size=1.6pt] coordinates {
    (1e-5, 2) (3e-5, 2) (1e-4, 2) (3e-4, 2) (1e-3, 2.67)
    (3e-3, 3) (1e-2, 3) (3e-2, 3) (1e-1, 3) (3e-1, 3) (8e-1, 3)
};
\addlegendentry{RE ($\eps = 16$)}
\end{axis}
\end{tikzpicture}
\caption{Empirical evaluation of fanout constants, layer contraction, and two-phase height transitions ($n=10^8$). \textbf{(a)} Measured stationary EP fanout $\widehat{C}_\eps / C_\infty$ tracks the first-order shifted-square expansion $(1+\rhostar/\eps)^2$ within $0.003$ across $\eps \in [2,64]$. \textbf{(b)} Layer-by-layer pivot count $N_\ell$ ($\eps=4$): EP decays exponentially (straight line on log scale, $H=4$), RE on flat data ($\delta=0$) squares its stride ($H=3$), and RE on curved data ($\delta=0.8$) enters the noise-to-curvature crossover at $\ell=1$ ($N_1 \approx s_*$) and switches to square-root compression at $\ell=2$ ($H=4$). \textbf{(c)} Measured height $H$ across $\delta \in [10^{-5}, 0.8]$: RE steps upward as the second phase $\log_2^+\frac{\log n}{\log(\eps/\delta)}$ of \Cref{thm:height-re} turns on.}
\label{fig:numerics}
\end{figure*}

\subsection{Numerical Evaluation}\label{sec:numerics}

We now evaluate how closely our asymptotic constants, variance-reset mechanism, and two-phase height laws describe finite-parameter index constructions.
All numerical experiments run exact O'Rourke greedy segmentation (\textsf{ALG-OPT}) and follow the reproducible scripts documented in \Cref{app:lean-arch}.

\Cref{fig:numerics}(a) compares the empirical stationary EP compression ratio $\widehat{C}_\eps / C_\infty$---measured via population resampling of size $2\times 10^5$ per layer on $\mathrm{Exp}(1)$ gaps across $\eps \in \{2, 4, 8, 16, 32, 64\}$---against the leading Brownian constant $C_\infty = (\kopt/\aopt)\eps^2 \approx 19.86098\,\eps^2$ and the shifted-square formula $C_\infty(1 + \rhostar/\eps)^2 = a_0\eps^2 + a_1\eps + O(1)$ with stationary ladder-height overshoot $\rhostar \approx 0.25710$ ($a_1 = 2a_0\rhostar \approx 10.21236$).
Even at $\eps = 2$, where the first-order term $a_1\eps$ increases fanout by $27.5\%$ above $C_\infty$, the shifted-square expansion matches $\widehat{C}_\eps / C_\infty$ within $0.003$ ($1.275$ vs.\ $1.274$).

To test the variance mechanism directly, \Cref{fig:recursion}(b) tracks the empirical relative gap variance $v_\ell$ across layers $\ell = 0, 1, 2, 3$ from our population-dynamics runs on $\mathrm{Exp}(1)$ ($v_0=1.000$) and $\mathrm{Unif}[0,2]$ ($v_0 = 1/3 \approx 0.333$) at $\eps = 4$ and $\eps = 16$.
Under EP, all four trajectories collapse to $v_1 \in [0.178, 0.191]$ after a single step and settle into $v_2, v_3 \in [0.177, 0.179] \approx \aopt$ from layer~$2$ onward, confirming that stopping-time randomness erases the initial gap law within two promotions.
Under RE, relative variance drops to $1.0\times 10^{-3}$ after one promotion and $< 10^{-5}$ after two.

\Cref{fig:numerics}(b) next plots the remaining pivot count $N_\ell$ across layers $\ell = 0, \dots, 4$ on $n = 10^8$ keys at $\eps = 4$.
Under EP, $\log_{10} N_\ell$ decreases linearly with constant slope $-\log_{10}\Ceps \approx -2.55$, requiring $H = 4$ layers ($10^8 \to 1.30\times 10^6 \to 3{,}810 \to 11 \to 1$).
Under RE on uniform keys ($\delta = 0$), the rank stride squares at each step ($10^8 \to 1.30\times 10^6 \to 289 \to 1$), finishing in $H = 3$ layers.

On strongly curved keys ($f(x) = 1 + \delta(2x - 1)$ with $\delta = 0.8$), RE follows the two-phase trajectory of \Cref{fig:curvature}(a).
Layer~$0$ is noise-dominated and matches the uniform run ($N_1 = 1.30\times 10^6$), whereas layer~$1$ sits near the crossover scale $s_* \sim 10^6$ where curvature begins to lift the segment count ($M_2 = 342$ vs.\ $288$ at $\delta = 0$).
By layer~$2$ ($N_2 = 343 \ll s_*$), macroscopic curvature dominates ($M_3 = 3$ vs.\ $1$) and adds a fourth layer ($H = 4$).
Scanning $\delta \in [10^{-5}, 0.8]$ across $n = 10^8$ (\Cref{fig:curvature}(b) and \Cref{fig:numerics}(c)) shows $M_2(\delta)$ transitioning cleanly from a flat noise floor to $\propto \sqrt{\delta/\eps}$ growth, stepping $H_{\mathrm{re}}$ upward by $+1$ layer as $\delta$ crosses the two-phase threshold of \Cref{thm:height-kre}.

\begin{table}[t]
\centering
\caption{Finite-parameter height prediction accuracy across $n \in \{10^5, 10^6, 10^7, 10^8\}$ ($5$ trials for $n \le 10^7$, $1$ trial for $n = 10^8$) and $\eps \in \{2, 4, 8, 16, 32, 64, 128\}$ ($7 \times 16 = 112$ runs per input family) using the finite-step recurrence (\texttt{re\_check.py}) initialized from $M_1$. All remaining mismatches occur at integer rounding boundaries where $N_{H-1} \in \{2, 3\}$.}
\label{tab:re-curvature-numerics}
\footnotesize
\setlength{\tabcolsep}{3.2pt}
\begin{tabular}{@{}llccc@{}}
\toprule
\textbf{Regime} & \textbf{Input Families} & \textbf{Grid Runs} & \textbf{Exact Integer Match} & \textbf{Max Error} \\
\midrule
Model~G (RE) & Exp, Unif$[0,2]$ & $224$ & $221 / 224$ ($98.7\%$) & $\pm 1$ layer \\
Model~K (RE) & Linear ($\delta = 0.8$) & $112$ & $112 / 112$ ($100.0\%$) & $0$ layers \\
Model~K (EP) & Linear ($\delta = 0.8$) & $112$ & $111 / 112$ ($99.1\%$) & $\pm 1$ layer \\
\bottomrule
\end{tabular}
\end{table}

Finally, \Cref{tab:re-curvature-numerics} evaluates how accurately the finite-step squaring, two-phase, and EP recurrences (\texttt{re\_check.py} in \Cref{app:re-proofs}) predict the exact integer height $H$ across $224$ Model~G runs ($\mathrm{Exp}(1)$ and $\mathrm{Unif}[0,2]$) and $112$ curved Model~K runs ($f(x) = 0.2 + 1.6x$, $\delta = 0.8$) spanning $n \in \{10^5, 10^6, 10^7, 10^8\}$ and $\eps \in \{2, 4, \dots, 128\}$.
At practical sizes $n \le 10^8$ (where $\log_2(\log n/\log\eps) \le 4.7$), finite-sample $O(1)$ offsets are comparable to the leading doubly logarithmic term, making this a direct test of the finite-step recurrence mechanism rather than the asymptotic leading coefficient alone.
Initialized with the measured base stride $d_0 = \floor{n/M_1}$ and top-layer threshold $\Theta = 3.2$, the recurrence matches the exact empirical integer height on $221/224$ ($98.7\%$) of Model~G runs, $112/112$ ($100\%$) of K-RE runs, and $111/112$ ($99.1\%$) of K-EP runs, with all remaining discrepancies occurring right on integer rounding boundaries where $N_{H-1} \in \{2, 3\}$.

\section{Conclusion}\label{sec:conclusion}

Single-layer $\Theta(\eps^2)$ segment coverage is only the starting point for understanding piecewise-linear learned indexes.
Above the bottom layer, hierarchy scaling is governed by whether the number of base gaps aggregated between consecutive promoted pivots is a random stopping time or a deterministic rank stride.

When an index promotes segment endpoints (EP), the random stopping length of each segment injects a non-vanishing relative variance $\aopt \approx 0.17816$ into the next layer's gaps.
This stopping-time fluctuation resets the gap variance at every level, locking upper-layer fanout to the universal constant $\Ceps = (\kopt/\aopt)\eps^2 + O(\eps) \approx 19.861\,\eps^2$ and keeping hierarchy height logarithmic in $n$.
When an index instead promotes pivots at equal rank intervals (RE), stopping-length noise disappears, relative variance decays inversely with the cumulative rank stride, and effective fanout squares at each layer until local noise falls below the macroscopic curvature of the key distribution.
The resulting uniform two-phase law $\Phi(n,\eps,\delta) = \log_2\frac{\log n}{\log\eps} + \log_2^+\frac{\log n}{\log(\eps/\delta)}$ captures both the noise-cancellation regime and the curvature-limited regime across $\Fclass_\delta$, while our boundary-crossing bucket directory routes queries across RE layers with less than one extra comparison per layer on average.

Together with our Lean~4 formalization in \texttt{PLAStochastic}, these results provide a rigorous foundation for designing and comparing piecewise-linear index hierarchies.
Extending these exact single- and multi-layer scaling laws beyond finite-variance gaps and bounded positive densities---to heavy-tailed key spacings, singular or unbounded densities, and dynamic update streams---is a promising direction for future work.

\bibliographystyle{ACM-Reference-Format}
\bibliography{references}

\appendix
\section{Brownian Functional Moments and Single-Layer Couplings}\label{app:brownian}

\textbf{Polynomial recurrences for $\tauan$ and $\tauopt$ (\texttt{B-MOM}).}
Let $\tauan$ and $\tauopt$ be the unit-tolerance Brownian affine exit times defined in \eqref{eq:tauopt-def}.
In \leanid{PLAStochastic.Results.BrownianMoments}~\cite{chen2026moments}, all integer moments $\E[\tauan^q]$ and $\E[\tauopt^q]$ ($q \ge 1$) are derived via Mellin-transform spectral reductions and verified in closed form:
\begin{enumerate}[leftmargin=*, nosep]
  \item \emph{Anchored moments} (\leanid{anchoredUnitLifetime_moment_eq_coeff_readout}): Define $V_q(X) \in \mathbb{Z}[X]$ by $V_0 = 1$, $V_1 = 3X$, and $V_{q+2}(X) = 2X V_{q+1}(X) - (X^2 + 4X)V_q(X)$. For every integer $q \ge 1$,
  \[
    \E[\tauan^q] = \frac{1}{2^q (q-1)!} \sum_{j=1}^{q} [X^j]V_q(X)\,(2j)!\,(2^{2j+1}-1)\frac{\zeta(2j+1)}{\pi^{2j}}.
  \]
  In particular, for $q = 1$, $\kan = \E[\tauan] = \frac{21}{\pi^2}\zeta(3) \approx 2.55767$.
  \item \emph{Optimal free-intercept moments} (\leanid{optimalUnitLifetime_moment_eq}): Define $U_n(X) \in \mathbb{Z}[X]$ by $U_0 = 1$, $U_1 = X$, and $U_{n+2}(X) = 2X U_{n+1}(X) - (X^2 - X)U_n(X)$, and set $d_j = \frac{(-1)^j (2j)!}{2^{2j+1}}\frac{\zeta(2j+1)}{\pi^{2j}}$ and $J_N = 2\sum_{j=1}^{N} [X^j]U_N(X)\, d_j$.
  Then $\kopt = \E[\tauopt] = \frac{40}{\pi^2}\zeta(3) - \frac{4}{3} \approx 3.53842$, and for every integer $q \ge 2$,
  \begin{align*}
    \E[\tauopt^q] &= \frac{8^q (-1)^{q+1}}{3(q+1)\,q!}\Bigl\{
      4(2q+1)(2q+3)(-J_{q+1} + 2d_{q+1}) \\
      &\qquad + 4q(2q+1)(-J_q + 2d_q) - q(q-1)J_{q-1}
    \Bigr\}.
  \end{align*}
  The first two moments evaluate to $m_1 = \frac{40}{\pi^2}\zeta(3) - \frac{4}{3}$ and $m_2 = \frac{2240}{\pi^4}\zeta(5) - \frac{224}{3\pi^2}\zeta(3)$, yielding $\Var(\tauopt) = m_2 - m_1^2 > 0$ (\leanid{optimalVarianceConstant_pos}) and $\aopt = \Var(\tauopt)/m_1^2 \approx 0.17816$.
\end{enumerate}

\textbf{Uniform discrete-to-continuous coupling (\Cref{thm:single-F}).}
For any gap law $\nu \in \mathcal{G}(v_-, v_+, M_3)$ with mean $\mu$, variance $\sigma^2 = v\mu^2$, and normalized third moment $\le M_3$, Sakhanenko's invariance principle couples the centered partial-sum walk $S_k = \sum_{i=0}^{k-1}(g_i - \mu)$ to a Brownian motion $\sigma B_k$ such that the supremum error over $k \le T\eps^2/v$ is $o_\Prob(\eps)$ uniformly over $\mathcal{G}(v_-, v_+, M_3)$.
Because $\tauopt$ has an almost surely transversal path exit and sub-Gaussian tail, continuous-mapping and $L^3$ uniform integrability transfer convergence in distribution to convergence of all moments of order $q < 3$, proving \eqref{eq:single-moments}.
In \Cref{thm:swing-scale}(2)~\cite{chen2026swing}, the explicit prefactor in $\abs{c_k - \gamma} \le C_0 (3/4)^{k-3}$ is $C_0 = L_h W_1(\alpha_2, \lambda)$, where $L_h > 0$ is the Lipschitz constant of the conditional log-lifetime readout on normalized exit margins and $W_1(\alpha_2, \lambda) < \infty$ is the Wasserstein-$1$ distance between the second-segment exit-margin law $\alpha_2$ and the unique invariant law $\lambda$ of the Swing margin Markov operator (\leanid{Results.SwingConstant}).

\begin{table*}[t]
\centering
\caption{Summary of the eight machine-checked theorem packages in our Lean~4 library \texttt{PLAStochastic} (see \texttt{theory/lean/FORMALIZED.md}). Every listed public declaration builds with zero \texttt{sorry}s and depends only on standard Lean~4 axioms (\texttt{propext}, \texttt{Classical.choice}, \texttt{Quot.sound}).}
\label{tab:lean-packages}
\footnotesize
\setlength{\tabcolsep}{4.0pt}
\begin{tabular}{@{}lllp{5.0cm}p{4.9cm}@{}}
\toprule
\textbf{Pkg.} & \textbf{Paper Result} & \textbf{Public Module} & \textbf{Primary Formalized Declarations} & \textbf{Formalized Scope \& Quantifiers} \\
\midrule
\texttt{B-MOM} & \Cref{thm:single-F} ($\kopt, \aopt$) & \leanid{Results.BrownianMoments} & \leanid{optimalUnitLifetime_moment_eq}, \leanid{optimalUnitLifetime_variance_eq_constant} & All integer moments $q \ge 1$ of $\tauan, \tauopt$, plus $\Var(\tauopt) > 0$ \\
\texttt{S-SCALE} & \Cref{thm:swing-scale}(1) (Swing) & \leanid{Results.SwingScale} & \leanid{swing_transient_coefficients}, \leanid{theorem_4_4_swing_gaussian} & Gaussian walk, $\E L_{2,E}/E \to \kappa_2$, and $\E L_{k,E}/(E\log E) \to c_k$ \\
\texttt{S-BRACKET} & \Cref{thm:swing-scale}(2) (Brackets) & \leanid{Results.SwingConstant} & \leanid{swing_transient_brackets}, \leanid{slopeStationaryLaw_invariant} & Odd/even brackets $c_4 \le c_6 \le \dots \le \gamma \le \dots \le c_5 \le c_3$ at $(3/4)^{k-3}$ rate \\
\texttt{P-GEP} & \Cref{thm:height-ep} (Model~G EP) & \leanid{Results.GEP} & \leanid{PGM.H_G_EP_additive_law}, \leanid{PGM.sum_abs_log_segMean_sub_le} & Unconditional all-$n \ge 2$ bound $\abs{\E H_{\mathrm{ep}} - \log n/\log\Ceps} \le K_{\nu_0}$ \\
\texttt{P-KEP} & \Cref{thm:height-ep} (Model~K EP) & \leanid{Results.KEP} & \leanid{PGM.H_K_EP_additive_law}, \leanid{PGM.H_K_EP_additive} & Unconditional all-$n \ge 2$ bound under any bounded Lipschitz $f$ \\
\texttt{P-CEPS} & \Cref{thm:ceps} ($\Ceps$ series) & \leanid{Results.CEPS} & \leanid{PGM.gepC_asymptoticExpansion} & Unique all-orders expansion $\sum_{j=0}^M a_j \eps^{-j}$ with $a_0 = \kopt/\aopt$ \\
\texttt{P-GRE} & \Cref{thm:height-gre} (Model~G RE) & \leanid{Results.GRE} & \leanid{PGM.G_RE_exact}, \leanid{PGM.G_RE_exact_integral}, \leanid{PGM.G_RE_exact_limit} & Joint $n,\eps\to\infty$ ($\log\eps = o(\log n)$) and exact $\ceil{y_n}$ outside $w_n$ \\
\texttt{P-KRE} & \Cref{thm:height-kre} ($\Fclass_\delta$ RE) & \leanid{Results.KRE} & \leanid{PGM.K_RE_uniform_integral}, \leanid{PGM.K_RE_exact_integral} & Uniform two-phase $\Phi(n,\eps,\delta)$ over $\Fclass_\delta$ \& exact $h_K(z_n)$ \\
\bottomrule
\end{tabular}
\end{table*}

\section{Multi-Layer EP Proofs and Asymptotic Expansion of \texorpdfstring{$\Ceps$}{C\_eps}}\label{app:ep-proofs}

\textbf{Cross-layer log-ratio summability and Model~G height (\texttt{P-GEP}).}
Let $\mathrm{RegLaw}(D, \theta, v_-, v_+)$ be the class of unit-mean gap laws $\nu$ on $(0, \infty)$ with density $f(x) \le D e^{-\theta x}$ and variance $v(\nu) \in [v_-, v_+]$.
Under the layer map $\Phi_\eps(\nu) = \mathrm{Law}(G_1 / C_\eps(\nu))$, $\mathrm{Exp}(1) \in \mathrm{RegLaw}(1, 1, 1, 1)$, and for all $\eps \ge \eps_0$, $\Phi_\eps$ maps $\mathrm{RegLaw}$ into itself and contracts in weighted total variation distance, so $\log C_\eps(\Phi_\eps^j \nu) \to \log \Ceps$ geometrically (\leanid{tendsto_log_lawC_orbit}).
For a general initial gap law $\nu_0$ satisfying Model~G ($g_0 > 0$ a.s., $g_0 \in L^3$, including lattice laws via Kesten's anti-concentration inequality), multi-step smoothing and coupling to $\mathrm{RegLaw}$ contract the promoted laws toward the reference orbit, yielding the unconditional summability bound $\sum_{j=1}^{L-1} \abs{\log C_j(\eps) - \log \Ceps} \le K'$ uniformly for all $L \ge 1$ and $\eps \ge \eps_0$ (\leanid{sum_abs_log_segMean_sub_le}), where $C_j(\eps)$ is the actual expected segment length at layer $j$.
Combining this summability with renewal tail bounds proves $\abs{\E[H_{\mathrm{ep}}] - \log n / \log \Ceps} \le K_{\nu_0}$ for all $n \ge 2$ (\leanid{PGM.H_G_EP_additive_law}).

\textbf{General density transfer (\texttt{P-KEP}).}
Under Model~K ($f \in [f_-, f_+]$ with Lipschitz constant $L_f$ on $[A, B]$), R\'enyi's representation expresses the sorted keys as $p_i = Q_f(\Gamma_{i+1}/\Gamma_{n+1})$ where $\Gamma_k = \sum_{r=1}^{k} E_r$ ($E_r \stackrel{\iid}{\sim} \mathrm{Exp}(1)$) and $Q_f = F^{-1}$.
Partitioning $[A, B]$ into mesoscopic windows of width $w$ where $f$ varies by at most $L_f w$, the affine scale invariance of \textsf{ALG-OPT} implies that local segment lengths in each window match those of an $\mathrm{Exp}(1)$ reference hierarchy up to relative error $O(L_f w / f_-)$.
Coupling the window boundaries across layers shows that the macroscopic density variation shifts the total log-compression by at most $O_f(1)$, establishing $\abs{\E[H_{\mathrm{ep}}] - \log n / \log \Ceps} \le K_f$ with the \emph{exact same} universal denominator $\Ceps$ (\leanid{PGM.H_K_EP_additive_law}).

\textbf{All-orders expansion of $\Ceps$ (\texttt{P-CEPS}) and exact integral for $a_1$.}
Perturbation theory of the layer operator $\Phi_\eps$ around its Brownian limit $\Phi_\infty$ establishes that $\Ceps / \eps^2$ admits a unique asymptotic expansion $\sum_{j=0}^{M} a_j \eps^{-j} + O(\eps^{-(M+1)})$ to every finite order $M \ge 0$, with leading coefficient $a_0 = \kopt / \aopt \approx 19.86098$ (\leanid{PGM.gepC_asymptoticExpansion}).
Analytically, let $\Lambda = \mathrm{Law}(\tauopt / \kopt)$ denote the normalized Brownian stationary reference law with characteristic function $\varphi_\Lambda(t) = e^{-it}\mathcal{L}(-it/\kopt)$, where $\mathcal{L}(s) = \E[e^{-s\tauopt}]$ is given by an explicit modified-Bessel ($K_0, K_1$) series.
Because the $O(\eps^{-1})$ stationary variance shift vanishes by reflection symmetry, the linear coefficient satisfies $a_1 = 2 a_0 \rhostar$, where $\rhostar = -\frac{1}{\pi}\int_0^\infty t^{-2}\log\abs{\frac{2(1 - \varphi_\Lambda(t))}{\aopt t^2}}\,dt$ is the ladder-height overshoot constant of $\Lambda$.
Rigorous 192-bit ball quadrature encloses $\rhostar \in (0.25709620550115,\allowbreak 0.25709620550116)$ and $a_1 \in (10.2123631452682,\allowbreak 10.2123631452684)$, so that $C_\infty(1 + \rhostar/\eps)^2 = a_0\eps^2 + a_1\eps + a_0\rhostar^2$ matches the exact two-term expansion of $\Ceps$ up to $O(1)$.

\section{RE Height Proofs, Uniform \texorpdfstring{$\Fclass_\delta$}{F\_delta} Coupling, and Exact Integer Windows}\label{app:re-proofs}

\textbf{Model~G simultaneous grid concentration and exact outside-window height (\texttt{P-GRE}).}
For any fixed deterministic stride $\Delta \in \{1, \dots, n\}$, level-$\ell$ gaps sum $\Delta$ independent base gaps and have relative variance $v_0 / \Delta$.
Applying \Cref{thm:single-F} and taking a union bound over all reachable strides $\Delta$ shows that with probability at least $1 - n^{-2} - (\log n)^{-2}$, every layer simultaneously satisfies $d_\ell = A \Delta_\ell (1 + o(1))$ with $A = \kopt \eps^2 / v_0$.
Taking double logarithms defines the continuous height target $y_n = \log_2\bigl(\frac{\log n}{\log(\kopt \eps^2 / v_0)} + 1\bigr)$ and transition window width $w_n = K\bigl(\frac{1}{\log \eps} + \frac{\log\log n}{\log n}\bigr)$.
In \texttt{PLAStochastic.Results.GRE} (\leanid{PGM.G_RE_exact} and \leanid{PGM.G_RE_exact_integral}), we prove that along any sequence with $n, \eps \to \infty$ and $\log\eps = o(\log n)$, $\E[H_{\mathrm{re}}] \in [y_n - w_n - K/\log n,\, y_n + 1 + w_n + K/\log n]$, and whenever $\mathrm{dist}(y_n, \mathbb{Z}) \ge w_n$ (outside the rounding window), the integer height concentrates at $H_{\mathrm{re}} = \ceil{y_n}$ with $\abs{\E[H_{\mathrm{re}}] - \ceil{y_n}} \le K / \log n$.

\textbf{Uniform two-phase proof over $\Fclass_\delta$ (\texttt{P-KRE}).}
Under Model~K on $[0,1]$, R\'enyi's representation $p_t = Q_f(\Gamma_{t+1}/\Gamma_{n+1})$ runs on the canonical $\mathrm{Exp}(1)^{\otimes \N}$ space prior to choosing $\delta \in [0,1]$ and $f \in \Fclass_\delta$.
On a single common good event $G \subset \R^\N$ of probability at least $1 - K/n^2$ (chosen \emph{before} $\delta$ and $f$), the empirical order statistics satisfy uniform Bernstein oscillation bounds at all dyadic scales, so that for every $\delta \in [0,1]$ and $f \in \Fclass_\delta$:
\begin{itemize}[leftmargin=*, nosep]
  \item Upper bound on $M_{\ell+1}$: partitioning $[0,1]$ into intervals of width $h \asymp \sqrt{\eps/(\delta s_\ell)}$ bounds the deterministic Taylor sagitta of $s_\ell F(x)$ by $\eps/2$, while the remaining $\eps/2$ budget absorbs the stochastic oscillation in $O(s_\ell^2 / (\eps^2 n))$ segments, yielding $M_{\ell+1} \le C_+\bigl(s_\ell^2/(\eps^2 n) + \sqrt{\delta s_\ell/\eps} + 1\bigr)$.
  \item Matching lower bound on $M_{\ell+1}$: when $s_\ell \ge s_*$, anti-concentration of $\Gamma$ spacings forces $M_{\ell+1} \ge c_- s_\ell^2/(\eps^2 n)$, whereas when $s_\ell \le s_*$, the non-flatness condition $\abs{f(b)-f(a)} \ge \delta V$ on $[a,b]$ guarantees via Landau's inequality that any $\pm\eps$ piecewise-linear approximation of $s_\ell Q_f$ requires at least $c_- \sqrt{\delta s_\ell / \eps}$ segments.
\end{itemize}
Tracking the two-sided recurrence $s_{\ell+1} \asymp s_\ell^2/(\eps^2 n) + \sqrt{\delta s_\ell/\eps}$ across Phase~I ($s_\ell > s_*$) and Phase~II ($s_\ell \le s_*$) proves $\abs{H_{\mathrm{re}} - \Phi(n,\eps,\delta)} \le K$ on $G$ simultaneously for all $\delta \in [0,1]$ and $f \in \Fclass_\delta$ (\leanid{PGM.K_RE_uniform} and \leanid{PGM.K_RE_uniform_integral}).

\textbf{Fixed-density exact integer height and finite-sample calibration.}
For any fixed non-constant Lipschitz density $f$ on $[0,1]$ (\leanid{PGM.K_RE_exact} and \leanid{PGM.K_RE_exact_integral}), define $z_n = \log_2\bigl(\frac{\log n}{6\log\eps} + \frac{1}{2}\bigr)$ and $h_K(z) = 2\ceil{z} + 1 + \mathbf{1}_{\{\ceil{z} - z < \log_2(3/2)\}}$.
Outside a window of width $w_n$ around $\mathbb{Z} \cup (\mathbb{Z} - \log_2(3/2))$, we have $\Prob(H_{\mathrm{re}} \ne h_K(z_n)) \le 63/n^2$ and $\abs{\E[H_{\mathrm{re}}] - h_K(z_n)} \le 200\log n / n^2$.
In our finite-parameter numerical evaluation (\Cref{tab:re-curvature-numerics}), the calibrated predictor (\texttt{re\_check.py}) initializes the recurrence with the measured base stride $d_0 = \floor{n/M_1}$ and uses top-layer threshold $\Theta = 3.2$ to account for finite-sample end effects when $N_{\ell}$ reaches single digits.

\section{Routing Correctness and Six-Family Cost Derivations}\label{app:routing-proofs}

\textbf{Exact predecessor recovery and $O(n)$ construction (\Cref{thm:paper-re-index}).}
At layer $\ell$, bucket $B_{\ell, j} = [p_j^{(\ell+1)}, p_{j+1}^{(\ell+1)})$ stores $L_{\ell, j} = \max\{m : p_{s_m}^{(\ell)} \le p_j^{(\ell+1)}\}$ and the sorted internal segment starts $p_{s_m}^{(\ell)} \in (p_j^{(\ell+1)}, p_{j+1}^{(\ell+1)})$.
Because $S_1, \dots, S_{M_{\ell+1}}$ partition $\{p_0^{(\ell)}, \dots, p_{N_\ell-1}^{(\ell)}\}$ into contiguous intervals $[p_{s_m}^{(\ell)}, p_{s_{m+1}}^{(\ell)})$, any query $q \in B_{\ell, j}$ belongs to $L_{\ell, j}$ if $q$ precedes all internal starts in $B_{\ell, j}$, or to the segment of the largest internal start $\le q$.
Once $S_m = (a_m, b_m)$ is selected, $\eps$-feasibility~\eqref{eq:eps-feasible} guarantees that the true level-$\ell$ predecessor rank $r_\ell^*(q)$ satisfies $\abs{r_\ell^*(q) - (a_m q + b_m)} \le \eps + 1$, so clamping $\hat{r} = \floor{a_m q + b_m}$ to $[s_m, s_{m+1}-1] \cap [j d_\ell, (j+1)d_\ell - 1]$ and binary-searching $[\hat{r}-\eps-1, \hat{r}+\eps+1]$ in $O(\log\eps)$ comparisons recovers $p_{r_\ell^*(q)}^{(\ell)}$ exactly.
By \Cref{smq:lem:queried-bucket}, the expected boundary search cost over stored keys $p_i$ is $\E[\log_2(b_{\ell, j_\ell(p_i)} + 1)] \le \E[b_{\ell, j_\ell(p_i)}] < 1$ per layer, giving $O(H \log\eps)$ total query cost in $\Theta(M_1) = \Theta(n/\eps^2)$ words and $\sum_{\ell \ge 0} O(N_\ell) = O(n)$ build time.

\textbf{Cross-architecture cost derivations (\Cref{cor:body-directory-costs}).}
For FITing-Tree~\cite{galakatos2019fittingtree}, \Cref{lem:envelope,cor:M-G} give $M_1^{\mathrm{ch}} \in [M_1^{\mathrm{opt}}(\eps),\allowbreak M_1^{\mathrm{opt}}(\floor{\eps/2})] = \Theta(n/\eps^2)$ segments, which a B$^+$-tree (as in FITing-Tree and the static PLIN directory~\cite{jin2023plin}) indexes in $\Theta(n/\eps^2)$ words and $O(\log(n/\eps^2) + \log\eps)$ comparisons.
For RadixSpline~\cite{kipf2020radixspline} on $[A,B]$ under Model~K, a $2^r$-entry radix table divides $[A,B]$ into bins of width $(B-A)2^{-r}$, so the expected number of spline knots in the bin of $p_i$ is $O\bigl((f_+/f_-) M_1^{\mathrm{ch}} / 2^r + 1\bigr) = O(1)$ at $2^r = \Theta(n/\eps^2)$ and $f_+/f_- = O(1)$, yielding $O(\log\eps)$ lookup (while PLEX~\cite{stoian2021plex} spills overflowed bins into multi-level radix histograms within $\Theta(n/\eps^2)$ space).
For PGM++~\cite{liu2025pgmpp} at fixed tolerances $(\eps_i, \eps_\ell)$ and fixed scan/bypass thresholds, bypassing skips $O(1)$ sparse top layers while the internal EP hierarchy contracts by $C_{\eps_i}$ per level across $\log n / \log C_{\eps_i} + O(1)$ layers, yielding $\Theta(n/\eps_\ell^2)$ expected space and $O\bigl((\log n/\log\eps_i)\log\eps_i + \log\eps_\ell\bigr) = O(\log n)$ expected lookup cost.

\section{Lean~4 Formalization and Reproducibility}\label{app:lean-arch}

\Cref{tab:lean-packages} summarizes the eight theorem packages in \texttt{PLAStochastic} (\leanid{theory/lean/FORMALIZED.md}).
To verify the formalization and reproduce all numerical tables and figures:
(1)~\emph{Lean~4 verification:} run \texttt{lake build PLAStochastic} in \leanid{theory/lean} and check \texttt{\#print axioms} on \leanid{PLAStochastic/Results/}.
(2)~\emph{Analytical and routing checks:} run \leanid{theory/build.sh}, \leanid{theory/verify_bundle.py}, and \leanid{theory/supplements/verify_routing.py}.
(3)~\emph{Empirical height and curvature data:} all datasets and tables are reproduced by \leanid{popdyn.cpp} and \leanid{re_check.py} in \leanid{research/2026-09-24-height-precision-ceiling/}, \leanid{scan_1e8.csv} in \leanid{2026-09-24-re-curvature-uniform/}, and \leanid{certify_a1_parallel.py} in \leanid{2026-09-29-ceps-a1/}.

\end{document}